\documentclass[paper=a4, fontsize=11pt]{scrartcl}
\usepackage[T1]{fontenc}
\usepackage{fourier}
\usepackage[english]{babel}															
\usepackage{amssymb}
\usepackage{amsmath,amsfonts,amsthm} 
\usepackage[pdftex]{graphicx}	
\usepackage{url}
\usepackage[title,titletoc,page]{appendix} 
\usepackage{listings} 
\usepackage{color}
\usepackage{cite}
\usepackage{verbatim}
\usepackage{sectsty}
\usepackage{bm}
\usepackage{subcaption}
\usepackage[ruled,vlined]{algorithm2e}
\allsectionsfont{\centering \normalfont\scshape}

\usepackage{fancyhdr}
\numberwithin{equation}{section}		
\numberwithin{figure}{section}			
\numberwithin{table}{section}				

\newtheorem{thm}{Theorem}[section]
\newtheorem{defi}[thm]{Definition}
\newtheorem{lem}[thm]{Lemma}

\newtheorem{prop}[thm]{Proposition}
\newtheorem{remark}{Remark}[section]
\newtheorem{assum}{Assumption}[section]

\newcommand{\vect}[1]{\boldsymbol{\mathbf{#1}}}
\newcommand{\babs}[1]{\Big|{#1}\Big|}
\newcommand{\btwonorm}[1]{\Big|\Big|{#1}\Big|\Big|_2}

\title{Mathematical Theory of Computational Resolution Limit in Multi-dimensions\\}
\author{
	Ping Liu\thanks{\footnotesize Department of Mathematics, 
		HKUST,  Clear Water Bay, Kowloon, Hong Kong (pliuah@connect.ust.hk).}
	\; and Hai Zhang\thanks{\footnotesize 
		Department of Mathematics, 
		HKUST,  Clear Water Bay, Kowloon, Hong Kong (haizhang@ust.hk). Hai Zhang was supported by HK RGC grant GRF 16304517 and GRF 16306318.}}

\date{}

\begin{document}
	\bibliographystyle{plain} 
	\maketitle
		
\begin{center}
\textbf{Abstract}
\end{center}
Resolving a linear combination of point sources from their band-limited Fourier data is a fundamental problem in imaging and signal processing. With the incomplete Fourier data and the inevitable noise in the measurement, there is a fundamental limit on the separation distance between point sources that can be resolved. This is the so-called resolution limit problem.  
Characterization of this resolution limit is still a long-standing puzzle despite the prevalent use of the classic Rayleigh limit. It is well-known that Rayleigh limit is heuristic and its drawbacks become prominent when dealing with data that is subjected to delicate processing, as is what modern computational imaging methods do. Therefore, more precise characterization of the resolution limit becomes increasingly necessary with the development of data processing methods. For this purpose, we developed a theory of ``computational resolution limit" for both number detection and support recovery in one dimension in \cite{liu2019computational, liu2020resolution}. 
In this paper, we extend the one dimensional theory to multi-dimensions.  More precisely, we define and quantitatively characterize the ``computational resolution limit" for the number detection and support recovery problems in a general $k$ dimensional space. Our results indicate that there exist a phase transition phenomenon regarding to the super-resolution factor and the signal-to-noise ratio in each of the two recovery problems. 
Our main results are derived using a subspace projection strategy.  Finally, to verify the theory, we proposed deterministic subspace projection based algorithms for the number detection and support recovery problems in dimension two and three. The numerical results confirm the phase transition phenomenon predicted by the theory.

\section{Introduction}

It is well-known that there is a fundamental diffraction limit in optical imaging systems due to the physical nature of wave propagation and diffraction. 
Since the first theory of diffraction limit by Ernst Abbe \cite{abbe1873beitrage, volkmann1966ernst}, there have been various limits or criterion proposed in the literature as candidates for the gauge of resolution limit, see for instance the Rayleigh limit \cite{rayleigh1879xxxi}. This initiated the long-standing debate of identifying the right resolution limit \cite{rayleigh1879xxxi, sparrow1916spectroscopic, born2013principles}, and the argument still goes on as new imaging technologies are constantly being developed \cite{den1997resolution, demmerle2015assessing}. We refer the readers to the appendix of \cite{chen2020algorithmic} for an excellent recount of history on the debates. On the other hand, from the perspective of mathematics, there is no resolution limit when one has perfect access of the exact intensity profile of the diffraction images. This simple fact is noticed by many \cite{di1955resolving, den1997resolution} and here we quote a remark of di Francia \cite{di1955resolving} for better exposition: `` Moreover it is only too obvious that from the mathematical standpoint, the image of
two points, however close to one another, is different from that of one point.'' Therefore, the resolution limit can only be rigorously set when taking into account the measurement noise or aberration to preclude perfect access to the diffraction images. As none of the classical resolution criterion is relevant to noise, they are mostly applicable for direct observation but not useful for data subjected to elaborate processing \cite{papoulis1979improvement, den1997resolution}. For example, in the case of two point sources, computer algorithms can be applied to discriminate the two sources to a smaller distance when the signal-to-noise ratio (SNR) is relatively high. Determining the number and exact position of two adjacent point sources then becomes a question of experimental precision dictated by photon statistics and noise level rather than being described by the Rayleigh limit. This kind of understanding motivates many researches for the two-point resolution from the perspective of statistical inference \cite{helstrom1964detection, helstrom1969detection, lucy1992statistical, lucy1992resolution, den1996model}. These attempts spanned the course of several decades in the last century,  see \cite{shahram2004imaging, shahram2004statistical, shahram2005resolvability} for a generalization and unification of the results in this direction. Therein, the authors derived explicit formula for the minimum SNR that is required to discriminate two point sources separated by a distance smaller than the Rayleigh limit. For the general case of $n$ point sources or infinity point sources, the first result was derived in 1992 \cite{donoho1992superresolution}. In recent years, due to the development of new super-resolution technologies \cite{schermelleh2010guide, huang2010breaking, cornea2014fluorescence} and sparsity-promoting super-resolution algorithms \cite{candes2014towards, tang2014near, duval2015exact}, there have been many works to investigate the resolving power of super-resolution algorithms, where a particular goal is to derive the minimax error estimation for the amplitude and support recovery. We refer the readers to \cite{demanet2015recoverability, batenkov2020conditioning, li2018stable, batenkov2019super} and the references therein for detail. Especially, the analysis of resolution limits of some popular super-resolution algorithms were presented in \cite{da2020stable, tang2015resolution, duval2015exact}.

Despite considerable research effort over the years, the mystery of resolution limit is still illusive. For instance, many established results only focus on the support and amplitude recovery with a prior information of source number. There is few result on the minimum resolvable distance of  $n$ point sources so that the source number can be detected exactly. To resolve the issue, we introduced the concept ``computational resolution limit'' for the number detection problem and quantitatively characterize it in dimension one in 
\cite{liu2019computational, liu2020resolution}. We also investigated the ``computational resolution limit'' for the support recovery in \cite{liu2020resolution}. We proved that the computational resolution limit for the number detection is of the order $\frac{1}{\Omega}(\frac{\sigma}{m_{\min}})^{\frac{1}{2n-2}}$, while for the support recovery is of the order $\frac{1}{\Omega}(\frac{\sigma}{m_{\min}})^{\frac{1}{2n-1}}$, where $\Omega$ is the cutoff frequency, $\sigma$ is the noise level and $m_{\min}$ is the minimum intensity of the sources. 
The goal of this paper is to extend these results from dimension one to multi-dimensions, where there is even fewer theoretical results despite the fact that many sophisticated multi-dimensional algorithms are in practical use, see for instance \cite{liao2015music, poon2019, karthikeyan2015formulation, zoltowski1996closed, wang2008tree, xi2014computationally, gu2015joint, wu2009doa, ye2009two, yilmazer2006matrix}. To our knowledge, only two theoretical results are available. In \cite{shahram2004statistical}, the authors derived explicit relationships between SNR and the minimum detectable distance of two point sources from their 2-dimensional image with zero-mean Gaussian white noise based on hypothesis testing. 
They showed that the SNR required to discriminate two point sources is inversely related to a polynomial of the separation distance of them. In \cite{chen2020algorithmic}, the authors formulated the resolution limit problem as a statistical inverse problem and, based on connections to provable algorithms for learning mixture models, they rigorously proved upper and lower bounds on the statistical and algorithmic complexity needed to resolve closely spaced point sources. Especially, they demonstrated that there is a phase transition where the sample complexity goes from polynomial to exponential. Similar phase transition of the required noise intensity for resolving 1-dimensional point sources is also reported in \cite{moitra:2015:SEF:2746539.2746561}.

In this paper, we investigate resolution limit for recovering a cluster of closely spaced point sources in a general $k$-dimensional space from their band-limited Fourier data.  We define and quantitatively characterize the computational resolution limit for the number detection problem. 
The characterization implies a phase transition phenomenon regarding to the super-resolution factor and the SNR in the detection of source numbers. 
Following a similar approach, we also define and quantitatively characterize the computational resolution limit for the support recovery problem. 
The characterization implies a similar phase transition phenomenon in the recovery of source supports.  These results are based on a subspace projection approach which 
reduces the $k$-dimensional problem to multiple $(k-1)$-dimensional problems. Such an approach was also used in  \cite{chen2020algorithmic}. 
Finally, to verify the theory, we propose subspace projection based algorithms for both number detection and support recovery. The numerical experiments in dimension two and three confirm the results on the phase transition phenomenon.

\subsection{Organization of the paper}
The rest of the paper is organized in the following way. 
In Section 2, we present the main results on computational resolution limit to which the proofs are provided in Section 3. In Section 4 and Section 5 we introduce respectively the subspace projection based number detection and support recovery algorithms in two and three dimensions. We also conduct numerical experiments which confirm the phase transition phenomenon. Finally, in Section 6, we present and prove some technical lemmas that are used in the subspace projection strategy.   

\section{Main results on computational resolution limits in multi-dimensions}\label{section:mainresults}
Throughout the paper, we consider the following model of a linear combination of point sources in a general k-dimensional space:
\[
\mu=\sum_{j=1}^{n}a_{j}\delta_{\vect y_j}
\]
where $\delta$ denotes Dirac's $\delta$-distribution in $\mathbb R^k$, $\vect y_j \in \mathbb R^k,1\leq j\leq n$, represent the support of the point sources and $a_j\in \mathbb C, j=1,\cdots,n$ their amplitudes. 
We call that measure $\mu$ is $n$-sparse if all $a_j$'s are not zero. 
We denote 
\[
m_{\min}=\min_{j=1,\cdots,n}|a_j|,
\quad 	d_{\min}=\min_{p\neq j}||\vect y_p-\vect y_j||_2.
\] 
We assume that the available measurement is 
\begin{equation}\label{equ:modelsetting1}
\mathbf Y(\vect{\omega}) = \mathcal F \mu (\vect{\omega}) + \mathbf W(\vect{\omega})= \sum_{j=1}^{n}a_j e^{i \vect{y}_j\cdot \vect{\omega}} + \mathbf W(\vect{\omega}), \ \vect \omega \in \mathbb R^k, \ ||\vect{\omega}||_2\leq \Omega,
\end{equation}
where $\mathcal F \mu$ denotes the Fourier transform of $\mu$, 
$\Omega$ is the cut-off frequency, and $\mathbf W$ is the noise. We assume that 
\[
 ||\mathbf W(\vect{\omega})||_\infty< \sigma,
\]
where $\sigma$ is the noise level. We are interested in the resolution limit for a cluster of tightly spaced point sources. 
To be more specific, we denote
\[
B_{\delta}^k(\vect x) := \Big\{ \mathbf y \ \Big|\ \mathbf y\in \mathbb R^k,\  ||\vect y||_2<\delta \Big\},
\] 
and assume that the following assumption holds. 
\begin{assum}\label{assum:problemsetup}\ \\
1: $\vect y_j \in B_{\frac{(n-1)\pi}{2\Omega}}^k(\vect 0)$ or $||\vect y_j||_2\leq \frac{(n-1)\pi}{2\Omega}$, $j=1,\cdots,n$;\\
2: $\vect Y(\vect \omega)= \mathcal F \mu(\vect \omega)+ \vect W(\vect \omega), ||\vect \omega||_2\leq \Omega$ with $||\vect W(\vect \omega)||_{\infty}<\sigma$.
\end{assum}

The inverse problem we are interested in is to recover the discrete measure $\mu$ from the above noisy measurement $\mathbf Y$. 
In the spacial domain, the inverse problem can be formulated as a convolution problem where the measurement is the convolution of point sources and a band-limited point spread function $f$. More precisely, in the presence of additive noise $\vect \epsilon(t)$, the measurement is 
\begin{equation} \label{eq-decon}
\vect y(\vect t)=\mu* f(\vect t)+\vect \epsilon(\vect t)=\sum_{j=1}^{n}a_j f(\vect t-\vect y_j)+\vect \epsilon(\vect t), \ \vect t\in \mathbb{R}^k.
\end{equation}
By taking Fourier transform on both sides, we obtain 
\begin{equation} \label{eq-deconv}
\mathcal{F} \vect y(\vect \omega)= \mathcal{F} f(\vect \omega)\cdot \mathcal F \mu(\vect \omega)+ \mathcal{F} \vect \epsilon(\vect \omega)= \mathcal{F}f(\vect \omega)(\sum_{j=1}^n a_j e^{i \vect y_j \cdot \vect \omega})+ \mathcal{F} \epsilon(\vect \omega),
\end{equation}
which is reduced to (\ref{equ:modelsetting1}).

\subsection{Computational Resolution Limit for number detection}
In this section, we introduce and characterize the computational resolution limit for the number detection problem in $k$-dimensions. Our main results are built upon delicate analysis of the $\sigma$-admissible measure defined below. 

\begin{defi}{\label{def:sigmaadmissiblemeasure}}
	Given measurement $\mathbf Y$, we say that $\hat \mu=\sum_{j=1}^{m} \hat a_j \delta_{ \mathbf{\hat y}_j}, \ \mathbf{\hat y}_j\in \mathbb R^k$ is a $\sigma$-admissible discrete measure of $\mathbf Y$ if
	\[
	||\mathcal F\hat \mu (\vect{\omega})-\vect Y(\vect{\omega})||_\infty< \sigma, \quad ||\vect{\omega}||_2\leq \Omega,\ \vect \omega \in \mathbb R^k.
	\]
\end{defi}

The set of $\sigma$-admissible measures of $\mathbf Y$ characterizes all possible solutions to the inverse problem with the given measurement $\mathbf Y$. A good reconstruction algorithm should give a $\sigma$-admissible measure. If there exists one $\sigma$-admissible measure with less than $n$ supports, then one may detect less than $n$ sources and miss the exact one if there is no additional a priori information. On the other hand, if all $\sigma$-admissible measures have at least $n$ supports, then one can determine the number $n$ correctly if one restricts to the sparsest admissible measures. 
This leads to the following definition of computational resolution limit to the number detection problem.


\begin{defi}\label{def:computresolutionlimitnumber}
The computational resolution limit to the number detection problem in $k$-dimensional space is defined as the smallest nonnegative number $\mathcal D_{k,num}$ such that for all $n$-sparse measure $\sum_{j=1}^{n}a_{j}\delta_{\mathbf y_j}, \vect y_j \in B_{\frac{(n-1)\pi}{2\Omega}}^{k}(\vect 0)$ and the associated measurement $\vect Y$ satisfying Assumption \ref{assum:problemsetup}, if 
	\[
	\min_{p\neq j} ||\mathbf y_j-\mathbf y_p||_2 \geq \mathcal D_{k, num},
	\]
then there does not exist any $\sigma$-admissible measure  with less than $n$ supports for $\mathbf Y$.
\end{defi}

The above resolution limit is termed ``computational resolution limit'' to  distinguish from the classic Rayleigh limit. Its explicit estimate in dimension one was presented in \cite{liu2020resolution}. Here we quantitatively characterize the computational resolution limit for the number detection problem in a general $k$-dimensional space. We denote
\begin{equation}\label{equ:defineofxi}
\xi(k)= \begin{cases}
\sum_{j=1}^{k}\frac{1}{j}, & \ k\geq 1,\\
0,& \ k=0.
\end{cases}
\end{equation}
We have the following upper bound for the computational resolution limit.

\begin{thm}\label{thm:highdupperboundnumberlimit0}
	Let the source $\mu =\sum_{j=1}^{n}a_j\delta_{\mathbf y_j}, \vect y_j \in \mathbb R^k$ and measurement $\mathbf Y$ satisfy the Assumption \ref{assum:problemsetup}. Let $n\geq 2$ and assume that the following separation condition is satisfied 
	\begin{equation}\label{equ:highdupperboundnumberlimit1}
	\min_{p\neq j, 1\leq p, j\leq n}\btwonorm{\mathbf y_p- \mathbf y_j}\geq \frac{4.4\pi e \ (\pi/2)^{s-1} (n(n-1)/\pi)^{\xi(s-1)}}{\Omega }\Big(\frac{\sigma}{m_{\min}}\Big)^{\frac{1}{2n-2}},
	\end{equation}
	where $\xi(\cdot)$ is defined in (\ref{equ:defineofxi}) and $s$ is the dimension of the smallest subspace in $\mathbb R^k$ which contains the set of points $\{\vect y_1, \cdots, \vect y_n \}$. Then there do not exist any $\sigma$-admissible measures of \,$\mathbf Y$ with less than $n$ supports.
\end{thm}

\begin{remark}
The constant factor before $\frac{\sigma}{m_{\min}}$ on the right hand side of (\ref{equ:highdupperboundnumberlimit1}) may not be optimal. Especially, the dependence on the source number $n$ is a result of the project strategy we used in the proof. 
We conjecture that the optimal constant is independent of the source number $n$. 
\end{remark}

\medskip
Compared with the Rayleigh limit $\frac{c_k\pi}{\Omega}$ where the constant $c_k$ depends on the spacial dimension $k$, Theorem \ref{thm:highdupperboundnumberlimit0} 
indicates that resolving the source number in the sub-Rayleigh regime is theoretically possible if the SNR is sufficiently large. We next show that the above upper bound is optimal in terms of the SNR.	

\begin{prop}\label{prop:highdnumberlowerbound0}
	For given $0<\sigma<m_{\min}$ and integer $n\geq 2$, there exist  $\mu=\sum_{j=1}^{n}a_j\delta_{\mathbf y_j}, \vect y_j \in \mathbb R^k$ with $n$ supports, and $\hat \mu=\sum_{j=1}^{n-1}\hat a_j \delta_{\mathbf {\hat y_j}}$ with $n-1$ supports such that 
	$||\mathcal F\hat \mu(\vect \omega)-\mathcal F \mu(\vect\omega)||_{\infty}< \sigma, ||\vect \omega||_2\leq \Omega$. Moreover
	\[
	\min_{1\leq j\leq n}|a_j|= m_{\min}, \quad \min_{p\neq j}\btwonorm{\mathbf y_p- \mathbf y_j}= \frac{0.81e^{-\frac{3}{2}}}{\Omega}\Big(\frac{\sigma}{m_{\min}}\Big)^{\frac{1}{2n-2}}.
	\]
\end{prop}

The above results indicate that
\[
\frac{C_{1,k}(n)}{\Omega}\Big(\frac{\sigma}{m_{\min}}\Big)^{\frac{1}{2n-2}} <  \mathcal D_{k,num} \leq \frac{C_{2,k}(n)}{\Omega }\Big(\frac{\sigma}{m_{\min}}\Big)^{\frac{1}{2n-2}}. 
\]
where $C_{1,k}(n)= 0.81e^{-\frac{3}{2}}$ and $C_{2,k}(n)=4.4\pi e (\pi/2)^{k-1} (n(n-1)/\pi)^{\xi(k-1)}$. 
The two bounds on the  computational  resolution  limit  for  number  detection imply an phase transition phenomenon in the number detection problem. 
Indeed, we define the super-resolution factor to be the ratio between Rayleigh limit and the minimum separation distance in the off-the-grid setting (the grid scale in the grid setting). In our case, since the Rayleigh limit is $\frac{c_k\pi}{\Omega}$ for some constant $c_k$ depending on space dimension, we define simply the super-resolution factor as 
\[
SRF:= \frac{\pi}{\Omega d_{\min}},
\]
where $d_{\min}=\min_{p\neq j}||\vect y_p-\vect y_j||_2$. 
We let  $SNR= \frac{m_{\min}}{\sigma}$ be the signal-to-noise ratio.  From the two bounds for the resolution limit, we can draw the conclusion that exact number detection is guaranteed if
\[
\log(SNR) > (2n-2)\log(SRF)+(2n-2) \log \frac{C_{1,k}(n)}{\pi}, 
\]
and may fail if
\[
\log(SNR) < (2n-2)\log(SRF)+(2n-2) \log \frac{C_{2,k}(n)}{\pi}.  
\]
It indicates that we expect two lines both of slope $2n-2$ in the parameter space of $\log SNR-\log SRF$ such that the number detection is successful for cases above the first line and unsuccessful for cases below the second. In the intermediate region between the two lines, the number detection can be either successful or unsuccessful from case to case. This is clearly demonstrated in the numerical experiments in dimension two and three in Section \ref{section:numberalgorithm}. 


\subsection{Computational Resolution Limit for stable support recovery}
We next present results on the resolution limit for the support recovery problem in $k$-dimensions. We first introduce the following concept of $\delta$-neighborhood of discrete measures. 
\begin{defi}\label{deltaneighborhood}
	Let  $\mu=\sum_{j=1}^{n}a_j \delta_{\vect y_j}, \vect y_j \in \mathbb R^k$ be a discrete measure and let $\delta>0$ be such that the $n$ balls $B_{\delta}^k(\vect y_j), 1\leq j \leq n$ are pairwise disjoint. We say that 
	$\hat \mu=\sum_{j=1}^{n}\hat a_{j}\delta_{\mathbf{\hat y}_j}$ is within $\delta$-neighborhood of $\mu$ if each $\mathbf {\hat y}_j$ is contained in one and only one of the $n$ balls $B_{\delta}^k(\vect y_j), 1\leq j \leq n$.
\end{defi}

According to the above definition, a measure in a $\delta$-neighborhood preserves the inner structure of the sources. For any stable support recovery algorithm, the output should be a measure in some $\delta$-neighborhood of the real measure. Moreover, $\delta$ should tend to zero as the noise level $\sigma$ tends to zero.  We now introduce the computational resolution limit for a stable support recovery. For ease of exposition, we only consider measures supported in $B_{\frac{(n-1)\pi}{2\Omega}}^{k}(\vect 0)$ where $n$ is the source number. 


\begin{defi}\label{def:computresolutionlimitsupport}
	The computational resolution limit to the stable support recovery problem is defined as the smallest non-negative number $\mathcal D_{k,supp}$ so that 
	for all $n$-sparse measure $\mu=\sum_{j=1}^{n}a_j \delta_{\mathbf y_j}, \vect y_j \in B_{\frac{(n-1)\pi}{2\Omega}}^{k}(\vect 0)$ and associated measurement $\vect Y$ satisfying Assumption \ref{assum:problemsetup}, 
	if
	\[
	\min_{p\neq j, 1\leq p,j \leq n} \btwonorm{\mathbf y_p-\mathbf y_j}\geq \mathcal{D}_{k,supp},
	\]  
	then there exists $\delta>0$ such that any $\sigma$-admissible measure for $\mathbf Y$ with $n$ supports in $B_{\frac{(n-1)\pi}{2\Omega}}^k(\mathbf 0)$ is within $\delta$-neighbourhood of $\mu$.  
\end{defi}

We have the following result on the characterization of $\mathcal{D}_{k,supp}$.

\begin{thm}\label{thm:highdupperboundsupportlimit0}
	Let $n\geq 2$, assume that $\mu=\sum_{j=1}^{n}a_j \delta_{\vect y_j}, \vect y_j \in \mathbb R^k$ and measurement $\vect Y$ satisfy Assumption \ref{assum:problemsetup} and the separation condition that 
	\begin{equation}\label{equ:highdsupportlimithm0equ0}
	d_{\min}:=\min_{p\neq j}\Big|\Big|\mathbf y_p-\mathbf y_j\Big|\Big|_2\geq \frac{5.88\pi e 4^{k-1}((n+2)(n-1)/2)^{\xi(k-1)}}{\Omega }\Big(\frac{\sigma}{m_{\min}}\Big)^{\frac{1}{2n-1}}, \end{equation}
	where $\xi(k-1)$ is defined as in (\ref{equ:defineofxi}). If $\hat \mu=\sum_{j=1}^{n}\hat a_{j}\delta_{\mathbf{\hat y}_j}$ supported on $B_{\frac{(n-1)\pi}{2\Omega}}^{k}(\vect 0)$ is a $\sigma$-admissible measure for $\vect Y$, then $\hat \mu$ is within the $\frac{d_{\min}}{2}$-neighborhood of $\mu$. Moreover, after reordering the $\mathbf{\hat y}_j$'s, we have 
	\begin{equation}\label{equ:highdsupportlimithm0equ1}
	\Big|\Big|\mathbf {\hat y}_j-\mathbf y_j\Big|\Big|_2\leq \frac{C(k, n)}{\Omega}SRF^{2n-2}\frac{\sigma}{m_{\min}}, \quad 1\leq j\leq n,
	\end{equation}
	where $C(k, n)=\big(4^{k-1}((n+2)(n-1)/2)^{\xi(k-1)}\big)^{(2n-1)}n2^{4n-2}e^{2n}\pi^{-\frac{1}{2}}$.
\end{thm}

\begin{remark}
The constant factor before $\frac{\sigma}{m_{\min}}$ on the right hand side of (\ref{equ:highdsupportlimithm0equ0}) may not be optimal. Especially, the dependence on the source number $n$ is a result of the project strategy we used in the proof. 
We conjecture that the optimal constant is independent of the source number $n$. Similar statement also holds for the constant factor $C(k, n)$. 
\end{remark}

Theorem \ref{thm:highdupperboundsupportlimit0} gives an upper bound for the computational resolution limit for stable support recovery in $k$-dimensional space. This bound is optimal in terms of the order of the SNR as is shown by the proposition below.   

\begin{prop}\label{prop:highdsupportlowerbound0}
	For given $0<\sigma<m_{\min}$ and integer $n\geq 2$, let 
	\begin{equation}\label{highdsupportlowerboundequ0}
	\tau=\frac{0.49e^{-\frac{3}{2}}}{\Omega}\ \Big(\frac{\sigma}{m_{\min}}\Big)^{\frac{1}{2n-1}}.
	\end{equation}
	Then there exist measure $\mu=\sum_{j=1}^{n}a_j \delta_{\vect y_j}, \vect y_j \in \mathbb R^k$ with $n$ supports at $\{(-\tau,0,\cdots,0), \cdots,  (-n\tau,0,\cdots, 0)\}$ and a measure $\hat \mu=\sum_{j=1}^{n}\hat a_j \delta_{\mathbf{\hat y}_j}$ with $n$ supports at  $\{(0,0,\cdots, 0),(\tau,0,\cdots, 0),\cdots, ((n-1)\tau, 0,\cdots, 0)\}$ such that
	$||\mathcal F \hat \mu(\vect \omega)-\mathcal F\mu(\vect \omega)||_{\infty}< \sigma, ||\vect \omega||_2\leq \Omega$
	and either $\min_{1\leq j\leq n}|a_j|= m_{\min}$ or $\min_{1\leq j\leq n}|\hat a_j|= m_{\min}$.   
\end{prop}

Proposition \ref{prop:highdsupportlowerbound0} provides a lower bound to the computational resolution limit $\mathcal{D}_{k,supp}$. Combined with Theorem \ref{thm:highdupperboundsupportlimit0}, it reveals that
\[
\frac{0.49e^{-\frac{3}{2}}}{\Omega}\ \Big(\frac{\sigma}{m_{\min}}\Big)^{\frac{1}{2n-1}}< \mathcal D_{k, supp} \leq \frac{5.88\pi e 4^{k-1}((n+2)(n-1)/2)^{\xi(k-1)}}{\Omega }\Big(\frac{\sigma}{m_{\min}}\Big)^{\frac{1}{2n-1}}.
\]

Similar to the number detection problem, the two bounds imply a phase transition phenomenon in the support recovery. More precisely, in the parameter space of $\log SNR-\log SRF$, we expect two lines both with slope $2n-1$ such that the support recovery is successful for cases above the first line and unsuccessful for cases below the second. In the intermediate region between the two lines, the support recovery can be either successful or unsuccessful from case to case. This is clearly demonstrated in the numerical experiments in dimension two and three in Section \ref{section:supportalgorithm}. 




\subsection{Remark on the projection based reconstruction approach}
We remark that the inverse problem of recovering point sources from their band-limited Fourier data is closely 
related to the problem of direction of arrival (DOA) which aims to determine the azimuth and elevation angles of a plane wave that impinge on an antenna array. The subspace projection based reconstruction approach was used in many classical DOA algorithms, see for instance \cite{van1992azimuth, yilmazer2006matrix, wang2008tree, wang2011computationally, xi2014computationally} where the azimuth and elevation angles are usually estimated separately and then paired together by some ad hoc schemes. Specifically, in these algorithms, one first recover the components in two or three specific one dimensional subspaces, typically the x, y, z axis. For such an projection approach to works it is required that the projected components in these subspaces are distinct.
In contrast, the subspaces are chosen in a random manner in \cite{chen2020algorithmic}. In this paper, we propose a strategy to 
choose the subspaces in a deterministic manner which can guarantee a stable recovery. The key idea follows from our theory that the stability of the support recovery depends crucially on the separation distance of the sources. Therefore 
the reconstructed projected components in the subspaces are more reliable when their minimum separation distance therein is larger. 
In practice, the multiple subspace measurements (sub-array measurements) can be obtained from the received data of the antenna arrays or by designing some specific array geometries which encompasses antennas aligned in multiple directions. 
On the other hand, in view of the development of computationally efficient high performance algorithms \cite{eriksson1994line, wang2011computationally} for recovering components in one dimensional space, the computational burden of recovering in multiple subspaces will not be a crucial issue in practical application. 
We expect our results can inspire new ideas for developing algorithms for multi-dimensional DOA problem.

\section{Proof of main results}\label{section:proofmainresults}
In this section, we prove the main results in the previous section using subspace projection approach and induction arguments. We first introduce some notations. For a vector $\vect v \in \mathbb R^k$, we denote $\vect v^\perp$ the orthogonal complement space of the 1-dimensional space spanned by $\vect v$. For $\vect y \in \mathbb R^k$ and a subspace $Q\subset \mathbb R^k$, we denote $\mathcal P_{Q}(\vect y)$ the orthogonal projection of $\vect y$ onto $Q$.

\subsection{Proofs of results for number detection}
\textbf{Proof of Theorem \ref{thm:highdupperboundnumberlimit0}:}\\
We first prove the case when the dimension of the smallest subspace containing all the point sources are exactly $k$, i.e. $s=k$. We  
prove the result by induction. We first 
note that the case $k=1$ is proved in Theorem 2.1 in \cite{liu2020resolution}. 
Suppose the claim holds for $k=l$, 
we validate the claim for $k = l+1$. By contradiction, assume that  for some $\mu=\sum_{j=1}^n a_j \delta_{\mathbf y_j},\ \vect y_j \in \mathbb R^{l+1}$ with measurements $\vect Y$ satisfying Assumption \ref{assum:problemsetup} and the following separation condition 
\begin{equation}\label{equ:proofhighdnumber1equ-1}
	\min_{p\neq j, 1\leq p, j\leq n}\btwonorm{\mathbf y_p- \mathbf y_j}\geq \frac{4.4\pi e(\pi/2)^{l} (n(n-1)/\pi)^{\xi(l)}}{\Omega }\Big(\frac{\sigma}{m_{\min}}\Big)^{\frac{1}{2n-2}}=: d_{\min},
\end{equation}
there exists a $\sigma$-admissible measure $\hat \mu =\sum_{j=1}^m \hat a_j \delta_{\mathbf {\hat y}_j}$ with $m<n$ such that
\begin{equation}\label{equ:proofhighdnumber1equ1}
||\mathcal{F} \hat \mu (\vect \omega) - \vect Y (\vect \omega)||_2<\sigma, \quad ||\vect \omega||_2\leq \Omega, \ \vect \omega\in \mathbb R^{l+1}.
\end{equation}
Let $\Delta=(\frac{\pi}{n(n-1)})^{\frac{1}{l}}$. By Lemma \ref{lem:highdnumberproject1}, there exists an unit vector $\vect v \in \mathbb R^{l+1}$ so that
\begin{equation}\label{equ:proofhighdnumber1equ0}
\min_{p\neq j, 1\leq p, j \leq n}\btwonorm{\mathcal P_{\vect v^\perp}(\vect{y}_p)-\mathcal P_{\vect v^\perp}(\vect{y}_j)}\geq \frac{2\Delta d_{\min}}{\pi}. 
\end{equation}
We now consider the discrete measure $\sum_{j=1}^{n}a_j \delta_{\mathcal P_{\vect v^\perp}(\mathbf{y}_j)}$ in the $l$-dimensional subspace $\vect v^{\perp}$ with measurement $\vect Y(\vect \omega), \vect \omega \in \vect v^{\perp}$. By (\ref{equ:proofhighdnumber1equ-1}) and (\ref{equ:proofhighdnumber1equ0}), we have 
\begin{equation}\label{equ:proofhighdnumber1equ2}
\min_{p\neq j, 1\leq p, j \leq n}\btwonorm{\mathcal P_{\vect v^\perp}(\vect{y}_p)-\mathcal P_{\vect v^\perp}(\vect{y}_j)}\geq \frac{4.4 \pi e (\pi/2)^{l-1} (n(n-1)/\pi)^{\xi(l-1)}}{\Omega}\Big(\frac{\sigma}{m_{\min}}\Big)^{\frac{1}{2n-2}}. 
\end{equation} 
Therefore, the separation condition (\ref{equ:highdupperboundnumberlimit1}) holds for the projected source locations $\mathcal P_{\vect v^\perp}(\mathbf{y}_j), j=1,\cdots,n$.  On the other hand, it is clear that $||\mathcal P_{\vect v^\perp}(\mathbf{y}_j)||_2\leq \frac{(n-1)\pi}{2\Omega}$. Hence Assumption \ref{assum:problemsetup} holds. 
Therefore, applying the claim of the theorem for  
dimension $k=l$, we can conclude that there should be no $\sigma$-admissible measures of \,$\mathbf Y(\vect \omega), \vect \omega \in \vect v^{\perp}$ with less than $n$ supports. However, (\ref{equ:proofhighdnumber1equ1}) implies
\begin{equation}\label{equ:proofhighdnumber1equ3}
\btwonorm{\sum_{j=1}^m \hat a_j e^{i\mathcal P_{\vect v^\perp}(\mathbf{\hat y}_j)\cdot \vect \omega} - (\sum_{j=1}^n \hat a_j e^{i\mathcal P_{\vect v^\perp}(\mathbf{y}_j)\cdot \vect \omega}+\vect W (\vect \omega))}<\sigma, \quad ||\vect \omega||_2\leq \Omega, \ \vect \omega \in \vect v^{\perp},
\end{equation}
which implies that $\sum_{j=1}^m \hat a_j \delta_{\mathcal P_{\vect v^\perp}(\mathbf{\hat y}_j)}$ is a $\sigma$ admissible measure of \,$\mathbf Y(\vect \omega), \vect \omega \in \vect v^{\perp}$. This is a  contradiction and it proves the claim for the case $k=l+1$. By induction, we have proved the claim for all $k\geq 1$. 

Finally, we prove the theorem for the case when $k>s$ where $s$ is the dimension of smallest subspace containing $\{\vect y_1, \cdots, \vect y_n \}$. We first find a $s$-dimensional subspace $S$ so that $\vect y_j \in S, 1\leq j \leq n$. By considering only the Fourier measurement in the space $S$, the problem is reduced to the one corresponding to the case $k=s$ which we proved previously.  This completes the proof of Theorem \ref{thm:highdupperboundnumberlimit0}.


\medskip
\textbf{Proof of Proposition \ref{prop:highdnumberlowerbound0}:}\\
Consider $\gamma =\sum_{j=1}^{2n-1} a_j \delta_{\vect t_j}$ with $\vect t_1 = (-(n-1)\tau, 0, \cdots, 0), \vect t_2 = (-(n-2)\tau, 0, \cdots, 0), \cdots, \vect t_{2n-1}= ((n-1)\tau, 0, \cdots, 0)$ and $\tau = \frac{0.81e^{-\frac{3}{2}}}{\Omega}\Big(\frac{\sigma}{m_{\min}}\Big)^{\frac{1}{2n-2}}$. For every $\vect \omega =(\omega_1, \omega_2, \cdots, \omega_k)^T$, $\mathcal F \gamma(\vect \omega) = \sum_{j=1}^{2n-1}a_je^{i \vect t_j \cdot \vect \omega} = \sum_{j=1}^{2n-1}a_je^{i (-n+j)\tau \omega_1}, |\omega_1|\leq \Omega$. This reduces the estimation of $\mathcal F \gamma(\vect \omega)$ to the one dimensional case. By Proposition 2.1 in \cite{liu2020resolution}, there exist $a_j, |a_j|\geq m_{\min}, 1\leq j\leq 2n-1$ so that $||\mathcal F \gamma(\vect \omega)||_{\infty}<\sigma$. As a consequence, 
\[
\mu = \sum_{j=1}^{n} a_j \delta_{\vect t_j}, \quad  \hat \mu  = \sum_{j=n+1}^{2n-1} -a_j \delta_{\vect t_j}
\]
yields the proposition.

\subsection{Proofs of results for support recovery}
\textbf{Proof of Theorem \ref{thm:highdupperboundsupportlimit0}:}\\
We prove the theorem by induction. We first note that the case when $k=1$ is exactly Theorem 2.2 in \cite{liu2020resolution}. 
Suppose that Theorem \ref{thm:highdupperboundsupportlimit0} holds for the case $k=l$, we now prove for the case $k=l+1$.  
Let $\mu=\sum_{j=1}^n a_j \delta_{\mathbf y_j}, \vect y_j \in \mathbb R^{l+1}$ and the associated measurement $\vect Y(\vect \omega), \vect \omega \in \mathbb R^{l+1}$ satisfy Assumption \ref{assum:problemsetup} and the minimum separation condition 
\begin{equation}\label{equ:highdsupportlimithm0equ-2}
	d_{\min}^{(l+1)}:=\min_{p\neq j, 1\leq p, j\leq n}\btwonorm{\mathbf y_p- \mathbf y_j} \geq \frac{5.88\pi e 4^{l} ((n+2)(n-1)/2)^{\xi(l)}}{\Omega }\Big(\frac{\sigma}{m_{\min}}\Big)^{\frac{1}{2n-1}},
\end{equation}
where $\xi(\cdot)$ is defined by (\ref{equ:defineofxi}). Assume $\hat \mu=\sum_{j=1}^n \hat a_j \delta_{\mathbf {\hat y}_j}, ||\mathbf {\hat y}_j||_2\leq \frac{(n-1)\pi}{2\Omega}, \mathbf {\hat y}_j \in \mathbb R^{l+1}$ is a $\sigma$-admissible measure. 
Let $\Delta = \frac{\pi}{8}(\frac{2}{(n+2)(n-1)})^{\frac{1}{l}}$. 
By Lemma \ref{lem:highdsupportproject1}, there exist $n+1$ unit vectors $\vect v_q$'s so that $0\leq \vect v_p \cdot \vect v_j \leq \cos 2\Delta, 1\leq p<j \leq n$, and for each $q$, 
\begin{equation}\label{equ:proofhighdsupport1equ0}
\min_{p\neq j}\btwonorm{\mathcal P_{\vect v_q^\perp}(\vect{y}_p)-\mathcal P_{\vect v_q^\perp}(\vect{y}_j)}\geq  d_{\min}^{(l)},
\end{equation}
where we define
\[
d_{\min}^{(l)} = \min_{p\neq j}\btwonorm{\vect y_p- \vect y_j} \frac{2\Delta}{\pi}=\frac{d_{\min}^{(l+1)}}{4((n+2)(n-1)/2)^{\frac{1}{l}}}.
\]
By (\ref{equ:highdsupportlimithm0equ-2})  we have
\begin{equation}\label{equ:proofhighdsupport1equ1}
\min_{p\neq j}\btwonorm{\mathcal P_{\vect v_q^\perp}(\vect{y}_p)-\mathcal P_{\vect v_q^\perp}(\vect{y}_j)}\geq d_{\min}^{(l)}> \frac{5.88\pi e 4^{l-1}((n+2)(n-1)/2)^{\xi(l-1)}}{\Omega}\Big(\frac{\sigma}{m_{\min}}\Big)^{\frac{1}{2n-1}}.
\end{equation}

Now for each $q$, consider the projected measure 
$\sum_{j=1}^n a_j \delta_{\mathcal P_{\vect v_q^\perp}(\mathbf{y}_j)}$ in the $l$-dimensional subspace $\vect v_q^\perp$ and the associated measurement $\vect Y(\vect \omega), \vect \omega \in \vect v_q^\perp$. It is clear that Assumption \ref{assum:problemsetup} and the separation condition (\ref{equ:highdsupportlimithm0equ0}) are satisfied. On the other hand, $\mathbf {\hat \mu}$ is a $\sigma$-admissible measure of $\vect Y$ implies $\sum_{j=1}^n \hat a_j \delta_{\mathcal P_{\vect v_q^\perp}(\mathbf{\hat y}_j)}$ is a $\sigma$-admissible measure of the measurement $\vect Y(\vect \omega), \vect \omega \in \vect v_q^{\perp}$. Using the assumption that Theorem \ref{thm:highdupperboundsupportlimit0} holds for the case $k=l$, we can conclude that for each $q$, we have a permutation $\tau_q$ of $\{1,\cdots, n\}$ so that
\begin{equation}\label{equ:proofhighdsupport1equ2}
\btwonorm{\mathcal P_{\vect v_q^\perp}(\mathbf{\hat y}_{\tau_q(j)})- \mathcal P_{\vect v_q^\perp}(\mathbf{y}_j)}\leq \frac{C(l, n)}{\Omega}\Big(\frac{\pi}{d_{\min}^{(l)}\Omega}\Big)^{2n-2}\frac{\sigma}{m_{\min}}, \quad 1\leq j\leq n.
\end{equation}

Note that for each fixed $\vect y_j$, we can find two different $q$'s, say, $q_1$ and $q_2$, such that $\mathbf{\hat y}_{\tau_{q_1}(j)} =\mathbf{\hat y}_{\tau_{q_2}(j)} = \mathbf{\hat y}_{p_j}$. 
Since $0\leq \vect v_{q_1} \cdot \vect v_{q_2} \leq \cos 2\Delta$,  we can apply Lemma \ref{lem:highdsupportproject2} to get
\[
\btwonorm{\mathbf {\hat y}_{p_j} - \vect y_j}\leq \frac{\sqrt{2}}{\sqrt{1-\cos(2\Delta)}} \frac{C(l, n)}{\Omega}\Big(\frac{\pi}{d_{\min}^{(l)}\Omega}\Big)^{2n-2}\frac{\sigma}{m_{\min}}, \quad 1\leq j\leq n.
\]

Using the inequality 
$1-\cos 2\Delta\geq \frac{8}{\pi^2}\Delta^2\geq \frac{1}{8}\big( \frac{2}{(n+2)(n-1)}\big)^{\frac{2}{l}}$, we further obtain
\begin{equation}\label{equ:proofhighdsupport1equ3}
\btwonorm{\mathbf {\hat y}_{p_j} - \vect y_j}\leq  \frac{4 ((n+2)(n-1)/2)^{\frac{1}{l}}C(l,n)}{\Omega}\Big(\frac{\pi}{d_{\min}^{l}\Omega}\Big)^{2n-2}\frac{\sigma}{m_{\min}}, \quad 1\leq j\leq n.
\end{equation}


We next claim that
\[
\btwonorm{\mathbf {\hat y}_{p_j} - \vect y_j}<\frac{d_{\min}^{(l+1)}}{2}.
\]
Indeed, by direct calculation, we can verify that
\[
4 ((n+2)(n-1)/2)^{\frac{1}{l}}C(l,n)\Big(\frac{1}{5.88e4^{l-1}((n+2)(n-1)/2)^{\xi(l-1)}}\Big)^{2n-2} < \frac{1}{2} 5.88\pi e  4^l((n+2)(n-1)/2)^{\xi(l)}.
\]
On the other hand, 
(\ref{equ:proofhighdsupport1equ1}) yields that
\[
(\frac{\pi}{d_{\min}^{(l)}\Omega})^{2n-2}\frac{\sigma}{m_{\min}}\leq \Big(\frac{1}{5.88e4^{l-1}((n+2)(n-1)/2)^{\xi(l-1)}}\Big)^{2n-2}\Big(\frac{\sigma}{m_{\min}}\Big)^{\frac{1}{2n-1}}.
\]
Therefore we have
\[
\frac{4((n+2)(n-1)/2)^{\frac{1}{l}}C(l,n)}{\Omega}\Big(\frac{\pi}{d_{\min}^{(l)}\Omega}\Big)^{2n-2}\frac{\sigma}{m_{\min}}< \frac{1}{2} \frac{5.88\pi e 4^l((n+2)(n-1)/2)^{\xi(l)}}{\Omega }\Big(\frac{\sigma}{m_{\min}}\Big)^{\frac{1}{2n-1}}. 
\]
The claim follows by combining the above inequality with (\ref{equ:proofhighdsupport1equ3}) and (\ref{equ:highdsupportlimithm0equ-2}).
So far, we have proved that for each $\vect y_j$, there exists a point 
$\mathbf {\hat y}_{p_j}$ which is in a $\frac{d_{\min}^{(l+1)}}{2}$ neighborhood of $\vect y_j$. Since these neighborhoods do not overlap, there exists only one $\mathbf {\hat y}_{p_j} \in \{\mathbf {\hat y}_{1},\cdots, \mathbf {\hat y}_{n}\}$ in the $\frac{d_{\min}^{(l+1)}}{2}$ neighborhood of $\vect y_j$. Thus we can reorder the index so that $\mathbf {\hat y}_{j}$ is in the $\frac{d_{\min}^{(l+1)}}{2}$ neighborhood of $\vect y_j$. Moreover we have
\[
\btwonorm{\mathbf {\hat y}_{j} - \vect y_j}\leq  \frac{(4((n+2)(n-1)/2)^{\frac{1}{l}})^{2n-1}C(l,n)}{\Omega}\Big(\frac{\pi}{d_{\min}^{(l+1)}\Omega}\Big)^{2n-2}\frac{\sigma}{m_{\min}}, \quad 1\leq j\leq n,
\]
which follows from (\ref{equ:proofhighdsupport1equ3}) and the equation that $d_{\min}^{(l)}=\frac{d_{\min}^{(l+1)}}{4((n+2)(n-1)/2)^{\frac{1}{l}}}$. This completes our induction argument and concludes the proof of Theorem \ref{thm:highdupperboundsupportlimit0}.

\medskip
\textbf{Proof of Proposition \ref{prop:highdsupportlowerbound0}:}\\
Similar to the proof of Proposition \ref{prop:highdnumberlowerbound0}, let $\gamma =\sum_{j=1}^{2n-1} a_j \delta_{\vect t_j}$ with $\vect t_j = (-(n+1-j)\tau, 0, \cdots, 0)$. The estimation of $\mathcal F \gamma (\vect \omega)$ can be reduced to one dimensional case. Employing the proof of Proposition 2.2 in \cite{liu2020resolution}, there exist $a_j, |a_j|\geq m_{\min}, 1\leq j\leq 2n-1$ so that $||\mathcal F \gamma (\vect \omega)||_{\infty}<\sigma$. Then
\[
\mu = \sum_{j=1}^{n} a_j \delta_{\vect t_j}, \quad  \hat \mu  = \sum_{j=n+1}^{2n} -a_j \delta_{\vect t_j},
\]
yields the desired result.

\vspace{0.5cm}
\section{Subspace projection based number detection algorithm}\label{section:numberalgorithm}
In this section, we propose a subspace projection based sweeping  singular-value-thresholding number detection algorithm in multi-dimensions based on Theorem \ref{thm:highdupperboundnumberlimit0}. 
For ease of exhibition, we only present the algorithm in dimension two and three.
We shall use the algorithm to demonstrate the phase transition phenomenons predicted by our theory in Section \ref{section:mainresults}. 


\subsection{1-dimensional sweeping singular-value-thresholding number detection algorithm}
In this section, we review the sweeping singular-value-thresholding number detection algorithm in 1 dimension \cite{liu2020resolution}. We refer to  \cite{akaike1998information, akaike1974new, wax1985detection, schwarz1978estimating, rissanen1978modeling, wax1989detection, lawley1956tests, chen1991detection, he2010detecting, han2013improved, liu2020resolution} and the references therein for other interesting algorithms in one dimension. 

For $\mu =\sum_{j=1}^n a_j \delta_{y_j}, y_j \in\mathbb R$ and measurement $\vect Y(\omega)$ satisfying Assumption \ref{assum:problemsetup}, we first choose a proper integer $s\geq n$ as an a prior estimation of the source number. We choose measurement at the sample points $z_t= -\Omega+\frac{t-1}{s}\Omega, t=1,\cdots,2s+1$:
\[
\mathbf Y(z_t)= \mathcal F \mu (z_t) + \mathbf W(z_t)= \sum_{j=1}^{n}a_j e^{i y_j z_t} +\mathbf W(z_t), \quad 1\leq t \leq 2s+1.
\]
We then form the following Hankel matrix
\begin{equation}\label{equ:hankelmatrix1}
\mathbf H(s)=\left(\begin{array}{cccc}
\mathbf Y(-\Omega)&\mathbf Y(-\Omega+\frac{1}{s}\Omega)&\cdots& \mathbf Y(0)\\
\mathbf Y(-\Omega+\frac{1}{s}\Omega)&\mathbf Y(-\Omega+\frac{2}{s}\Omega)&\cdots&\mathbf Y(\frac{1}{s}\Omega)\\
\cdots&\cdots&\ddots&\cdots\\
\mathbf Y(0)&\mathbf Y(\frac{1}{s}\Omega)&\cdots&\mathbf Y(\Omega)
\end{array}
\right),
\end{equation}
and consider the singular value decomposition of $\mathbf H(s)$ 
\[\mathbf H(s)=\hat U\hat \Sigma \hat U^*,\]
where $\hat\Sigma =\text{diag}(\hat \sigma_1,\cdots, \hat \sigma_n, \hat \sigma_{n+1},\cdots,\hat\sigma_{s+1})$ with the singular values $\hat \sigma_j$, $1\leq j \leq s+1$, ordered in a decreasing manner. We then determine the source number by thresholding on these singular values with a properly chosen threshold based on Theorem 5.1 in \cite{liu2020resolution}. The procedure is summarized in \textbf{Algorithm \ref{algo:onedsimplenumber}} below. Note that in \textbf{Algorithm \ref{algo:onedsimplenumber}}, it is required that the input integer $s$ is greater than the source number $n$. However, a suitable $s$ is not easy to estimate and large $s$ may incur a deterioration of resolution. To remedy the issue, we proposed a sweeping singular-value-thresholding number detection algorithm which allows us to find the minimum (or sparsest) source number from those admissible measures; see \textbf{Algorithm \ref{algo:onedsweepingnumber}} below.

\begin{algorithm}[H]\label{algo:onedsimplenumber}
	\caption{\textbf{Singular-value-thresholding number detection algorithm}}
	\textbf{Input:} Number $s$, Noise level $\sigma$\\
	\textbf{Input:} measurement: $\mathbf{Y}=(\mathbf Y(\omega_1),\cdots, \mathbf Y(\omega_M))^T$\\	
	1: $r=(M-1)\mod 2s$,  $\mathbf{Y}_{new}=(\mathbf Y(\omega_1), \mathbf Y(\omega_{r+1}), \cdots, \mathbf Y(\omega_{2sr+1}))^T$\;
	2: Formulate the $(s+1)\times(s+1)$ Hankel matrix $\mathbf H(s)$ from $\mathbf{Y}_{new}$, and
	compute the singular value of $\mathbf H(s)$ as $\hat \sigma_{1}, \cdots,\hat \sigma_{s+1}$ distributed in a decreasing manner\;
	4: Determine $n$ by $\hat \sigma_n>(s+1)\sigma$ and $\hat \sigma_{j}\leq (s+1)\sigma, j=n+1,\cdots, s+1$\;
	\textbf{Return:} $n$ 
\end{algorithm}

\begin{algorithm}[H]\label{algo:onedsweepingnumber}
	\caption{\textbf{Sweeping singular-value-thresholding number detection algorithm}}	
	\textbf{Input:} Noise level $\sigma$, measurement: $\mathbf{Y}=(\mathbf Y(\omega_1),\cdots, \mathbf Y(\omega_M))^T$\\
	\textbf{Input:} $n_{max}=0$\\
	\For{$s=1: \lfloor \frac{M-1}{2}\rfloor$}{
		Input $s,\sigma, \mathbf{Y}$ to \textbf{Algorithm 1}, save the output of \textbf{Algorithm 1} as $n_{recover}$\; 
		\If{$n_{recover}>n_{max}$}{$n_{max}=n_{recover}$}
	}
	{
		\textbf{Return} $n_{max}$
	}
\end{algorithm}

\subsection{Multi-dimensional sweeping singular-value-thresholding number detection algorithm}\label{section:highdnumberalgorithm}
We now derive the multi-dimensional sweeping singular-value-thresholding number detection algorithms. The strategy is to detect the source number in some properly chosen low-dimensional subspace. 

We first look at the 2-dimensional case. 
To be specific, let $\mu=\sum_{j=1}^{n}a_j \delta_{\vect y_j}, \vect y_j\in \mathbb R^2$ and let $\vect Y(\vect \omega), ||\vect \omega||_2\leq \Omega$ be the associated measurement.  We first choose the following $\frac{n(n+1)}{2}$ unit vectors
\begin{equation}
\vect v(\theta_q)=(\cos{\theta_q}, \sin\theta_q)^T\in \mathbb R^2, \ q=1,\cdots,\frac{n(n+1)}{2},
\end{equation}
where $\theta_q= \frac{q2\pi}{n(n+1)}$. For each $q$, we form Hankel matrix $\vect H_{q}(s)$ in the same way as (\ref{equ:hankelmatrix1}) from the measurement in the subspace $\vect v(\theta_q)^{\perp}$. Denote $\hat \sigma_{q,j}$ the $j$th singular value of $\vect H_{q}(s)$, we can detect the exact source number by thresholding on $\hat \sigma_{q,j}$'s under a suitable separation condition as is shown in the theorem below.

\begin{thm}\label{thm:highdMUSICthm1}
Let $n\geq 2, s\geq n$ and $\mu=\sum_{j=1}^{n}a_j \delta_{\vect y_j}, \vect y_j\in \mathbb R^2$ with $\vect y_j\in B_{\frac{(n-1)\pi}{2\Omega}}^k(\vect 0), 1\leq j\leq n$. For the singular values of $\ \vect H_{q}(s)$, we have 
\begin{equation}\label{equ:highdMUSICthm1equ-1}
\hat \sigma_{q,j}\leq  (s+1)\sigma, \quad j=n+1,\cdots,s+1, \quad q=1,\cdots, \frac{n(n+1)}{2}.
\end{equation}
Moreover, if the following separation condition is satisfied
\begin{equation}\label{equ:highdMUSICthm1equ0}
d_{\min}:=\min_{p\neq j}\btwonorm{\vect y_p-\vect y_j}>\frac{\pi sn(n+1)}{2\Omega}\Big(\frac{n(s+1)}{\zeta(n)^2}\frac{\sigma}{m_{\min}}\Big)^{\frac{1}{2n-2}},
\end{equation}
there exists $q^*$ so that 
\begin{equation}\label{equ:highdMUSICthm1equ2}
\hat\sigma_{q^*,n}>(s+1)\sigma.
\end{equation}
\end{thm}
Proof: Note that for each $q$ the projected measure $\sum_{j=1}^{n}a_j \delta_{\mathcal P_{\vect v(\theta_q)^{\perp}}(\vect y_j)}$ on the 
subspace $\vect v(\theta_q)^\perp$
satisfies Assumption \ref{assum:problemsetup}. By applying Theorem 5.1 in \cite{liu2020resolution} to the Hankel matrix $\vect H_{q}(s)$ formulated from the measurement $\vect Y(\vect \omega)$ where $\vect \omega$ is restricted to the subspace $\vect{v}(\theta_q)^{\perp}$,  we get (\ref{equ:highdMUSICthm1equ-1}) immediately. Moreover, when separation condition (\ref{equ:highdMUSICthm1equ0}) holds, by Lemma \ref{lem:highdmusicdirection}, there is some $q^*$ so that 
\[
\min_{p\neq j, 1\leq p, j \leq n}\babs{\mathcal P_{\vect v(\theta_{q^*})^\perp}(\vect{y}_p)-\mathcal P_{\vect v(\theta_{q^*})^\perp}(\vect{y}_j)}\geq \frac{2 d_{\min}}{n(n+1)}> \frac{\pi s}{\Omega}\Big(\frac{2n(s+1)}{\zeta(n)^2}\frac{\sigma}{m_{\min}}\Big)^{\frac{1}{2n-2}}.
\]
Applying Theorem 5.1 in \cite{liu2020resolution} again, we get $\hat\sigma_{q^*,n}>(s+1)\sigma$.

\medskip
The above theorem shows that for point sources that are well separated, we 
can determine the correct source number $n$ by 
thresholding on the singular values of the Hankel matrices $\vect H_q(s)$'s.
We note that the number of required unit vectors $\vect v(\theta_q)$ is not available since $n$ is unknown. In practice, we can choose a large enough $N$, say $N\geq \frac{n(n+1)}{2}$.
We summarize our algorithm as below.

\begin{algorithm}[H]\label{algo:twodsweepingnumber}
	\caption{\textbf{Two-dimensional sweeping singular-value-thresholding number detection algorithm}}	
	\textbf{Input:} Noise level $\sigma$, measurement: $\mathbf{Y}(\vect \omega), \vect \omega \in \mathbb R^2, ||\vect \omega||_2\leq \Omega$, and $n_{\max} = 0$\\
	\textbf{Input:} A large enough $N$, and corresponding $N$ unit vectors $\vect v(\theta_q), \theta_q = \frac{q\pi}{N}, q=1,\cdots, N$ \\
	\For{$q=1,\cdots,N$}{
		Input $\sigma$ and $\mathbf{Y}(\vect \omega), \vect \omega \in \vect v(\theta_q)^{\perp}$ to \textbf{Algorithm 2}, save the output of \textbf{Algorithm 2} as $n_{recover}$\; 
		\If{$n_{recover}>n_{max}$}{$n_{max}=n_{recover}$}
	}
	{
		\textbf{Return} $n_{max}$
	}
\end{algorithm}

\medskip
We now consider the number detection algorithm in $3$-dimensions. Similar to Theorem \ref{thm:highdMUSICthm1}, when sources are well separated, we can recover the exact source number $n$ by applying \textbf{Algorithm \ref{algo:twodsweepingnumber}} to measurement in some properly chosen 2-dimensional subspace. Precisely, for $N=\lfloor (\frac{n(n-1)}{2})^{\frac{1}{2}}\rfloor+1$, we denote the following unit vectors  
\begin{equation}
\vect v(\phi_1, \phi_{2})=(\cos\phi_1 \sin \phi_2, \ \sin {\phi_1}\sin {\phi_2}, \ \cos \phi_2)^T, \quad \phi_1, \phi_2\in \Big\{ \frac{\pi}{2N}, \frac{2\pi}{2N}, \cdots, \frac{\pi}{2}\Big\}.
\end{equation}
Then we use the measurement in each of the 2-dimensional subspaces $\vect v(\phi_1, \phi_2)^{\perp}$'s and utilize \textbf{Algorithm 3} to detect the source number therein. 
The algorithm is summarized as below.

\begin{algorithm}[H]\label{algo:threedsweepingnumber}
	\caption{\textbf{Three-dimensional sweeping singular-value-thresholding number detection algorithm}}	
	\textbf{Input:} Noise level $\sigma$, measurement: $\mathbf{Y}(\vect \omega), \vect \omega \in \mathbb R^3, ||\vect \omega||_2\leq \Omega$, and $n_{\max} = 0$\\
	\textbf{Input:} A large enough $N$, and corresponding $N^2$ unit vectors $\vect v(\phi_1, \phi_{2})=(\cos\phi_1 \sin \phi_2, \ \sin {\phi_1}\sin {\phi_2}, \ \cos \phi_2)^T, \quad \phi_1, \phi_2\in \Big\{ \frac{\pi}{2N}, \frac{2\pi}{2N}, \cdots, \frac{\pi}{2}\Big\}$.\\
	\For{$\phi_1, \phi_2\in \big\{ \frac{\pi}{2N}, \frac{2\pi}{2N}, \cdots, \frac{\pi}{2}\big\}$}{
		Input $\sigma$ and $\mathbf{Y}(\vect \omega), \vect \omega \in \vect v(\phi_1, \phi_2)^\perp$ to \textbf{Algorithm 3} (with some modifications of the $\vect v(\theta_q)$'s therein), save the output of \textbf{Algorithm 3} as $n_{recover}$\; 
		\If{$n_{recover}>n_{max}$}{$n_{max}=n_{recover}$}
	}
	{
		\textbf{Return} $n_{max}$
	}
\end{algorithm}

\subsection{Numerical experiments and phase transition}
In this section we conduct numerical experiments to demonstrate the phase transition phenomenon regarding to the super-resolution factor and the SNR using \textbf{Algorithm \ref{algo:twodsweepingnumber}} and \textbf{\ref{algo:threedsweepingnumber}}. 
In view of the computational resolution limit for the number detect (which is of the order $O(\frac{1}{\Omega}\big(\frac{\sigma}{m_{\min}}\big)^{\frac{1}{2n-2}})$), recovering $5$ closely-spaced point sources demands extremely low noise level. Therefore, in the experiments we only consider the case with three or four point sources. 
For the $2$-dimensional case, we fix $\Omega =1$ and detect the source number from their noisy Fourier measurement. The noise intensity is $\sigma$ and the minimum separation distance of point source is $d_{\min}$. We perform $10000$ random experiments (the randomness is in the choice of ($d_{\min}$,$\sigma$, $\vect y_j$, $a_j$)) and detect the number by \textbf{Algorithm \ref{algo:twodsweepingnumber}}. For the 3-dimensional case, We detect the source number by \textbf{Algorithm \ref{algo:threedsweepingnumber}} and perform 10000 random experiments under the same setup as the 2-dimensional case.

\begin{figure}[!h]
	\centering
	\begin{subfigure}[b]{0.28\textwidth}
		\centering
		\includegraphics[width=\textwidth]{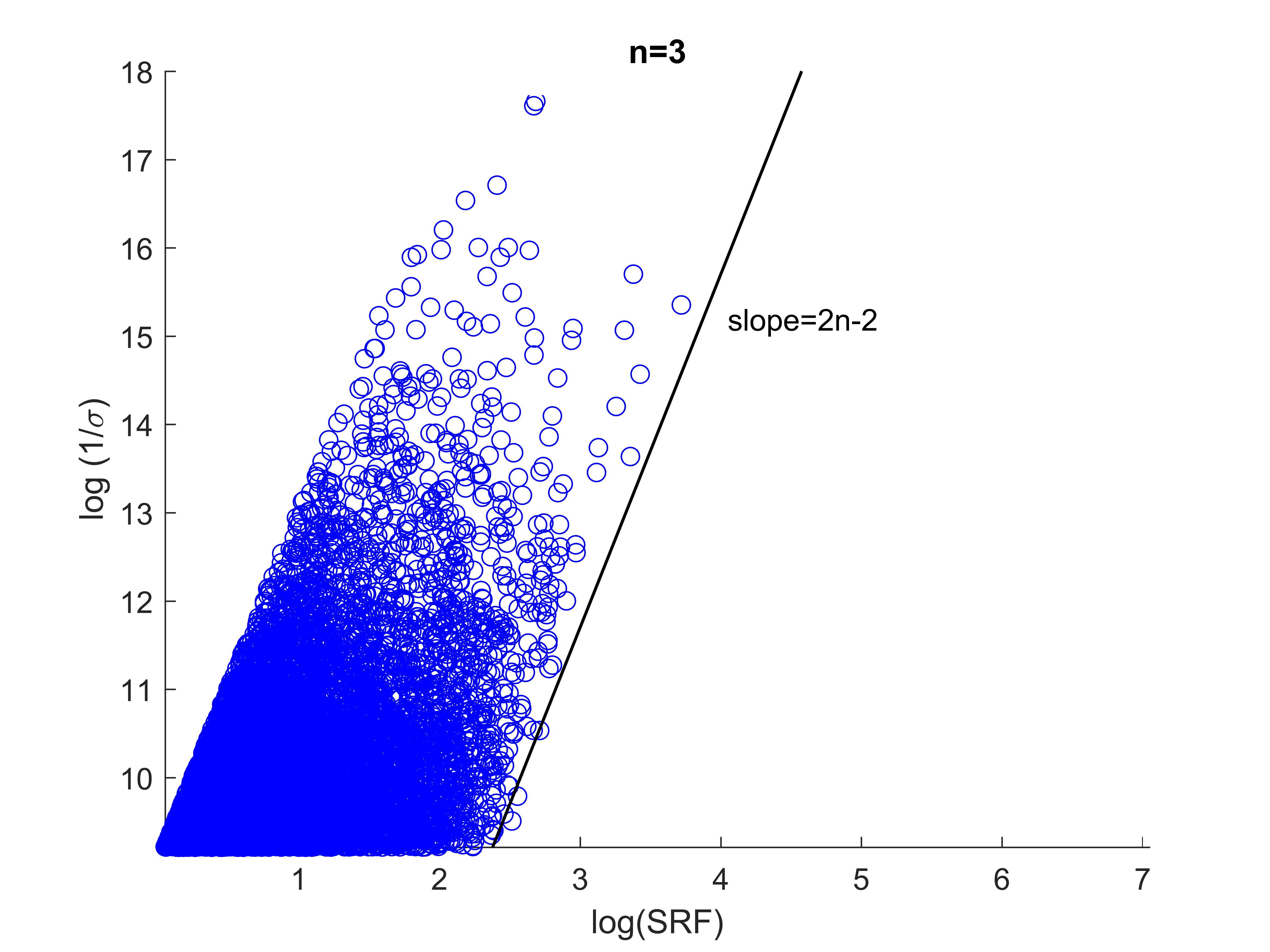}
		\caption{detection success}
	\end{subfigure}
	\begin{subfigure}[b]{0.28\textwidth}
		\centering
		\includegraphics[width=\textwidth]{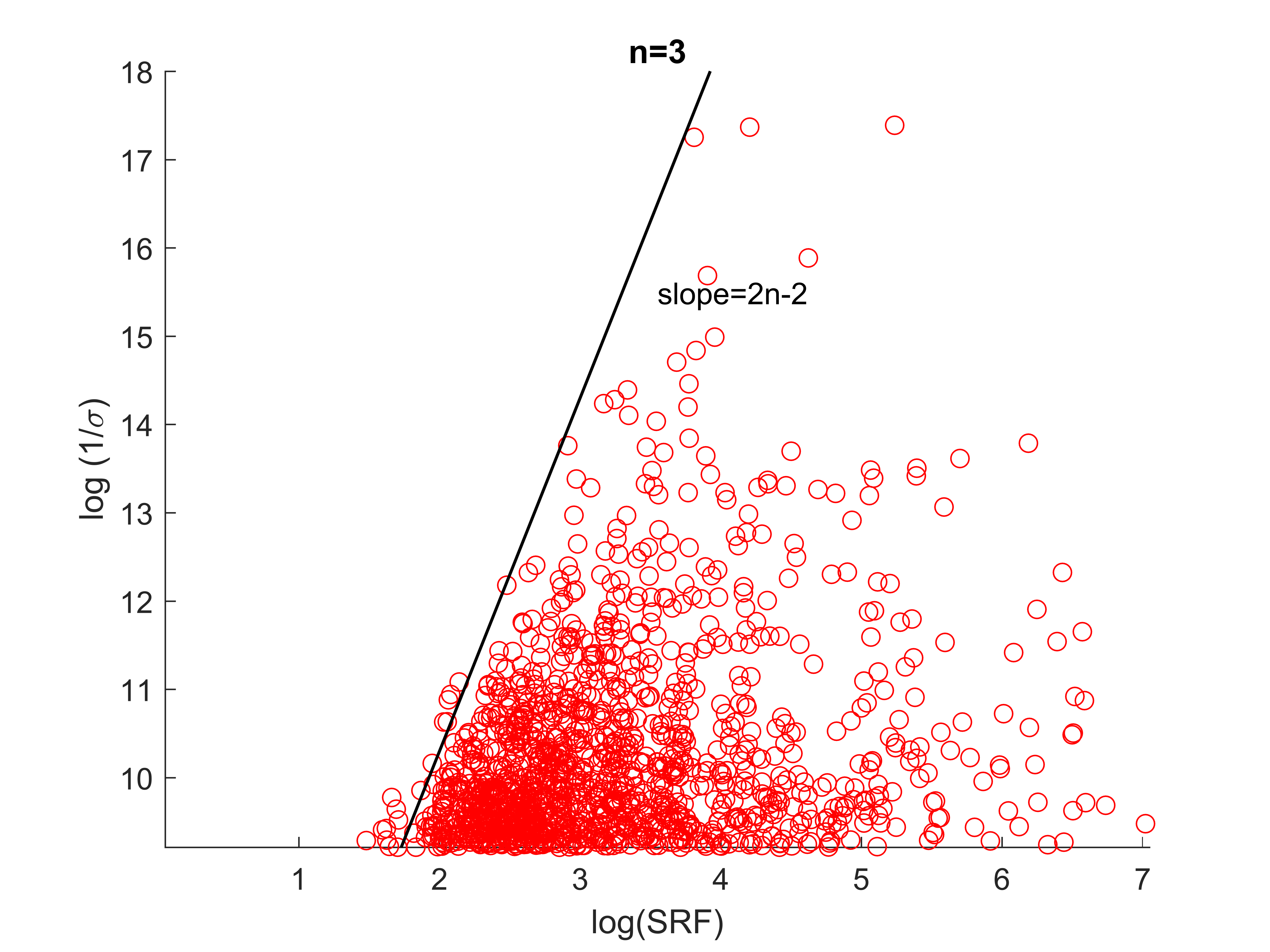}
		\caption{detection fail}
	\end{subfigure}
	\begin{subfigure}[b]{0.28\textwidth}
		\centering
		\includegraphics[width=\textwidth]{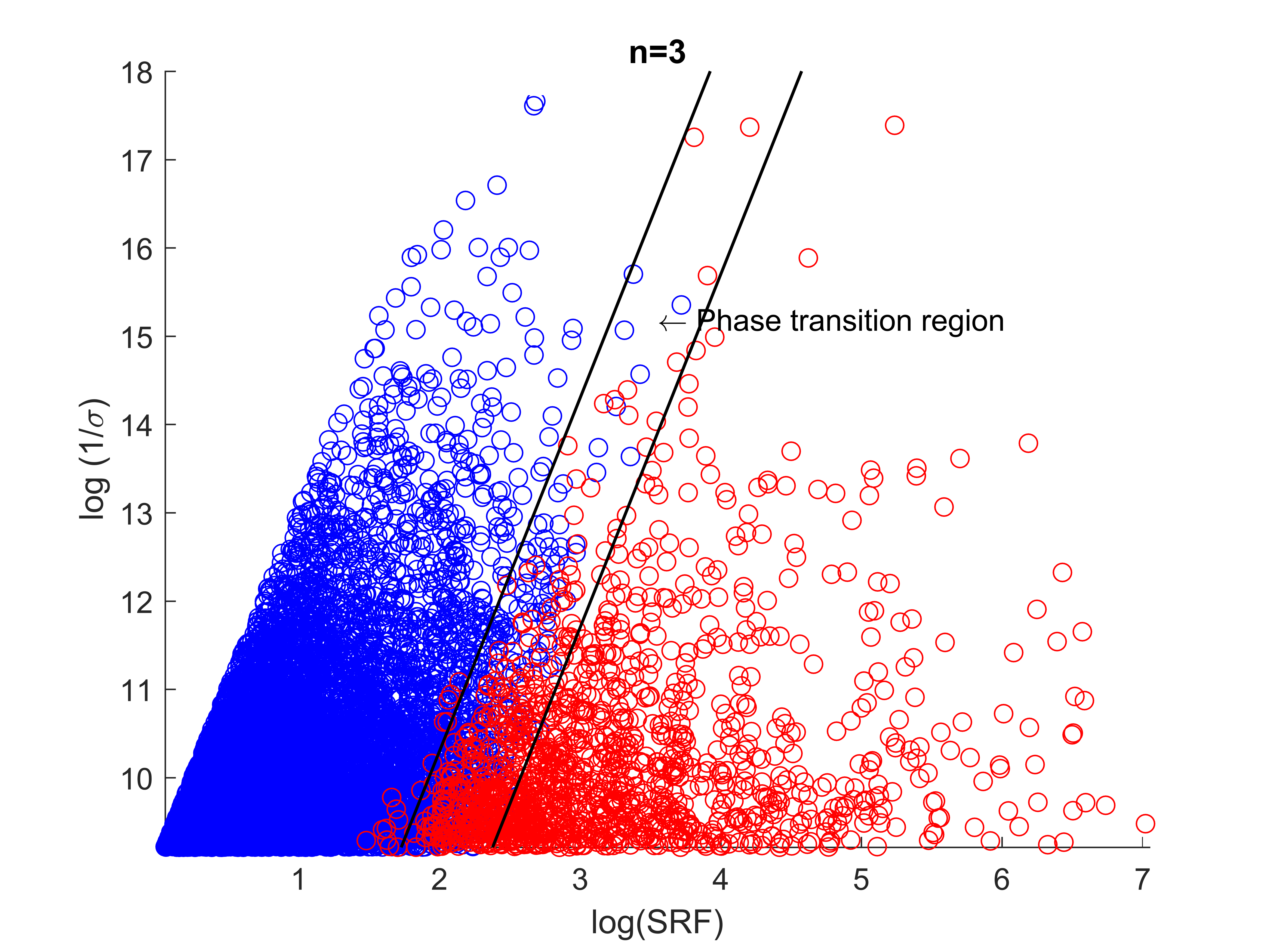}
		\caption{phase transition region}
	\end{subfigure}
	\begin{subfigure}[b]{0.28\textwidth}
		\centering
		\includegraphics[width=\textwidth]{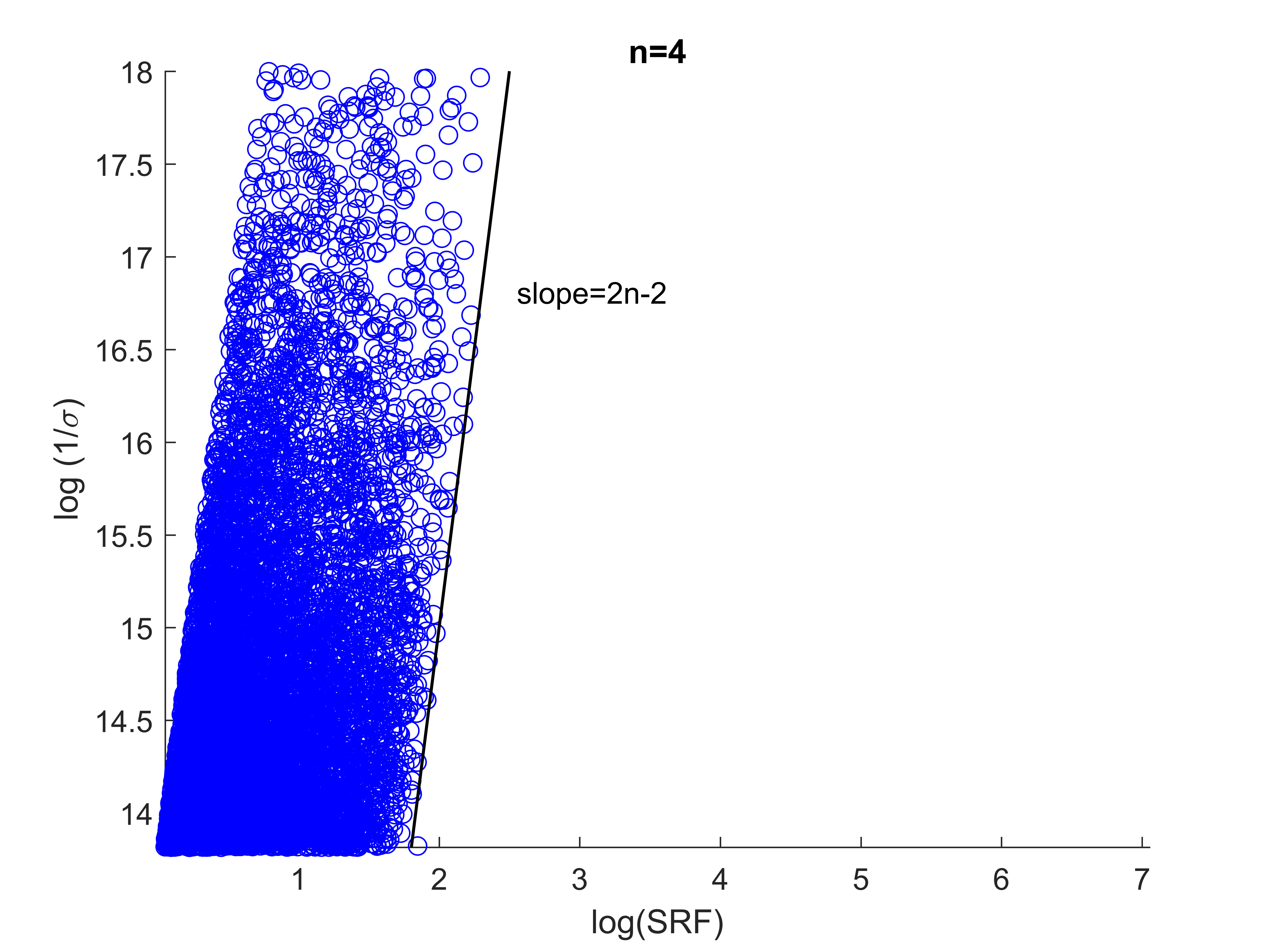}
		\caption{detection success}
	\end{subfigure}
	\begin{subfigure}[b]{0.28\textwidth}
		\centering
		\includegraphics[width=\textwidth]{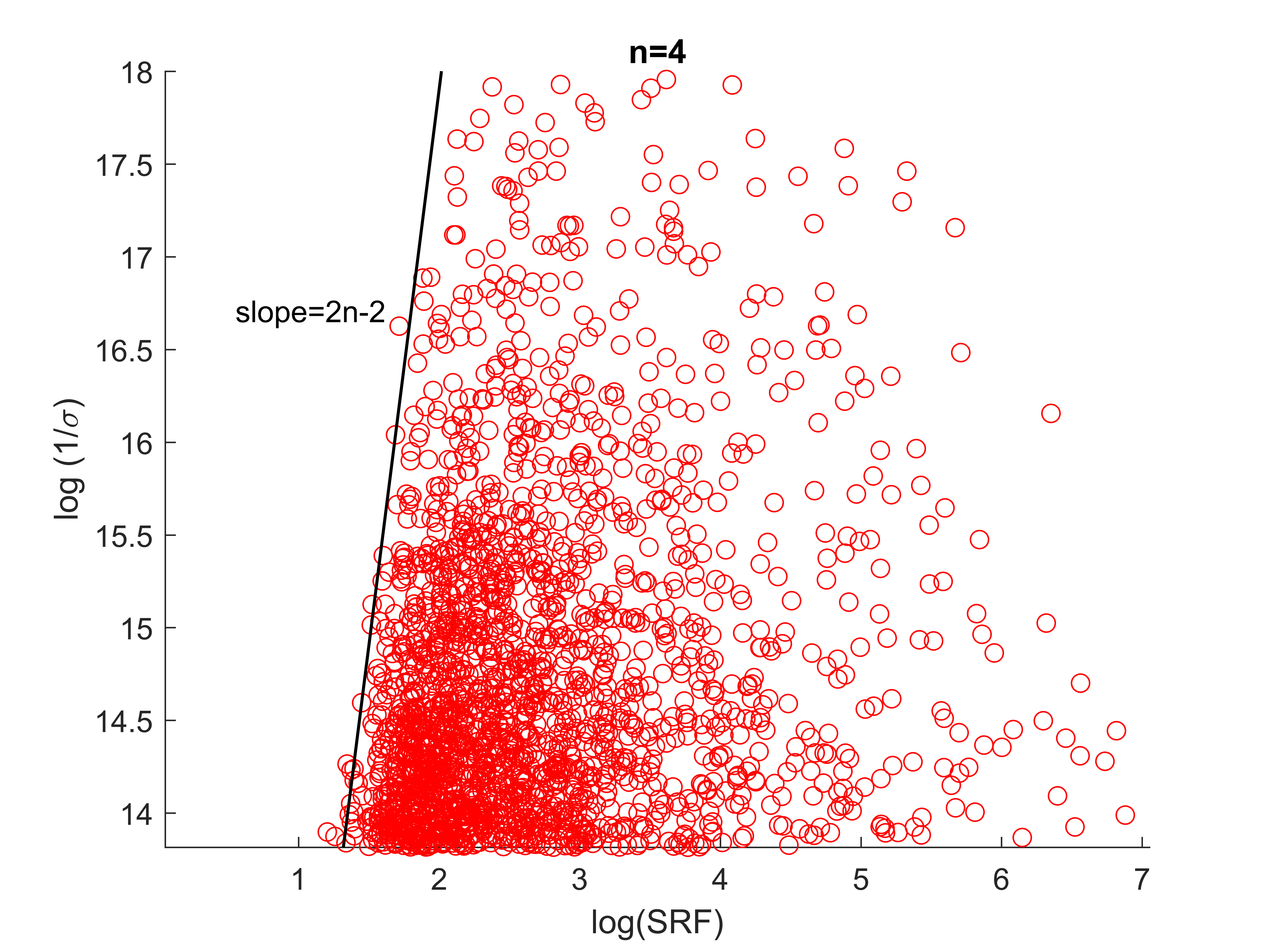}
		\caption{detection fail}
	\end{subfigure}
	\begin{subfigure}[b]{0.28\textwidth}
		\centering
		\includegraphics[width=\textwidth]{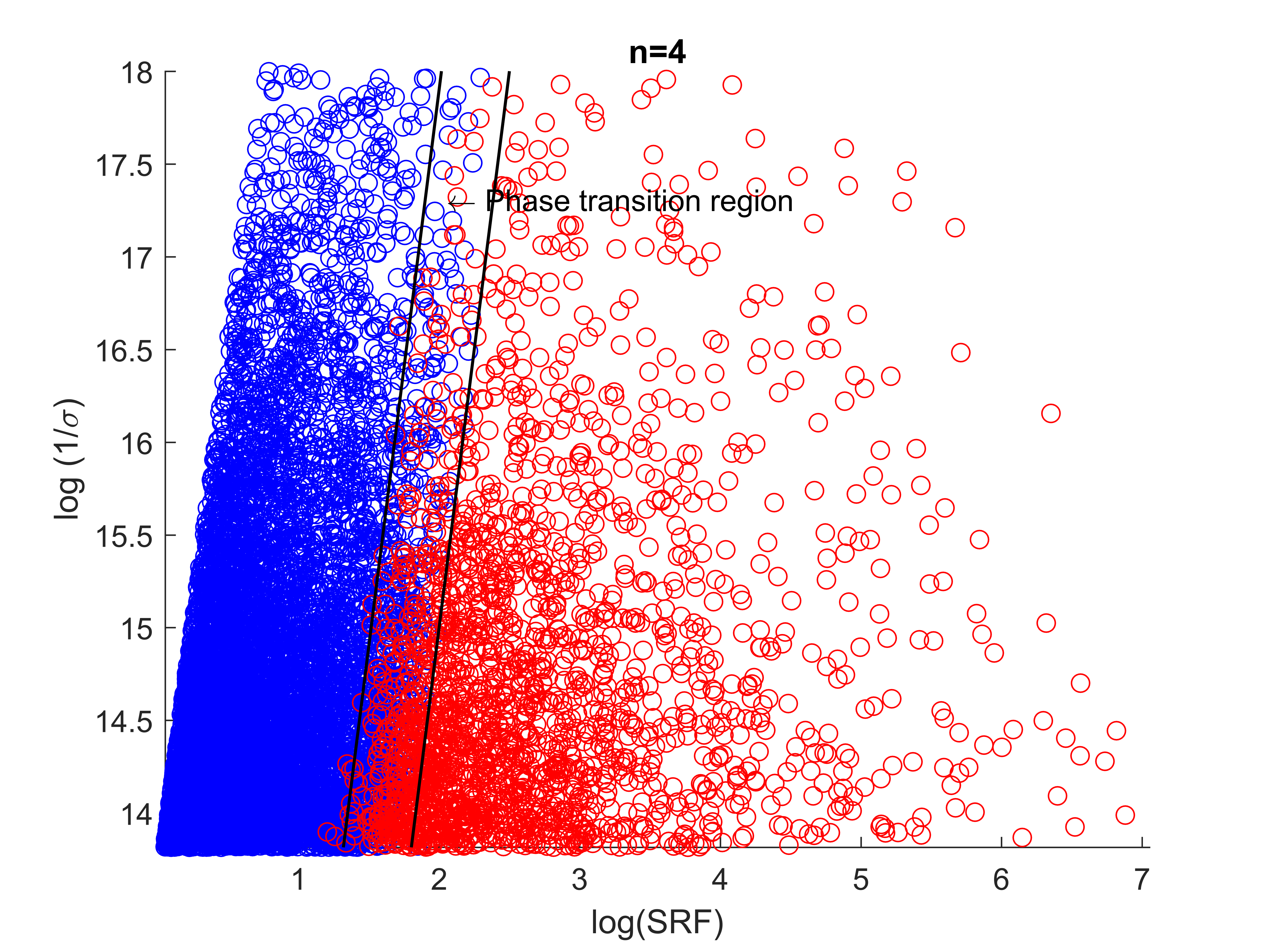}
		\caption{phase transition region}
	\end{subfigure}
	\caption{Plots of the successful and the unsuccessful number detection by \textbf{Algorithm \ref{algo:twodsweepingnumber}} depending on the relation between $\log(SRF)$ and $\log(\frac{1}{\sigma})$. (a) illustrates that three point sources can be exactly detected if $\log(\frac{1}{\sigma})$ is above a line of slope $4$ in the parameter space. Conversely, for the same case, (b) shows that the number detection fails if $\log(\frac{1}{\sigma})$ falls below another line of slope $4$. (f) highlights the phase transition region which is bounded by the black slashes in (a) and (b). (d),(e) and (f) illustrate parallel results for four point sources.}
	\label{fig:twodnumberphasetransition}
\end{figure}
\begin{figure}[!h]
	\centering
	\begin{subfigure}[b]{0.28\textwidth}
		\centering
		\includegraphics[width=\textwidth]{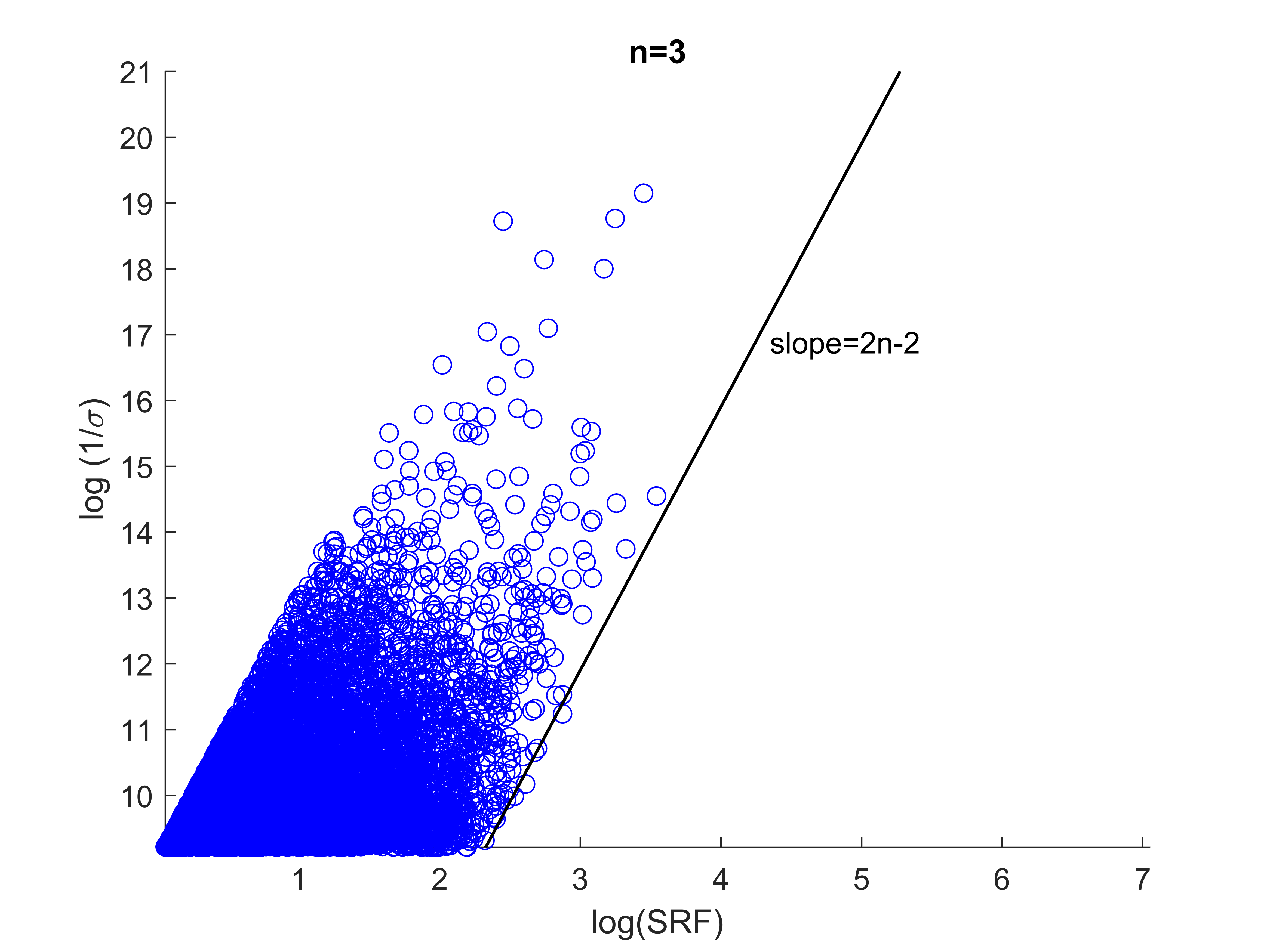}
		\caption{detection success}
	\end{subfigure}
	\begin{subfigure}[b]{0.28\textwidth}
		\centering
		\includegraphics[width=\textwidth]{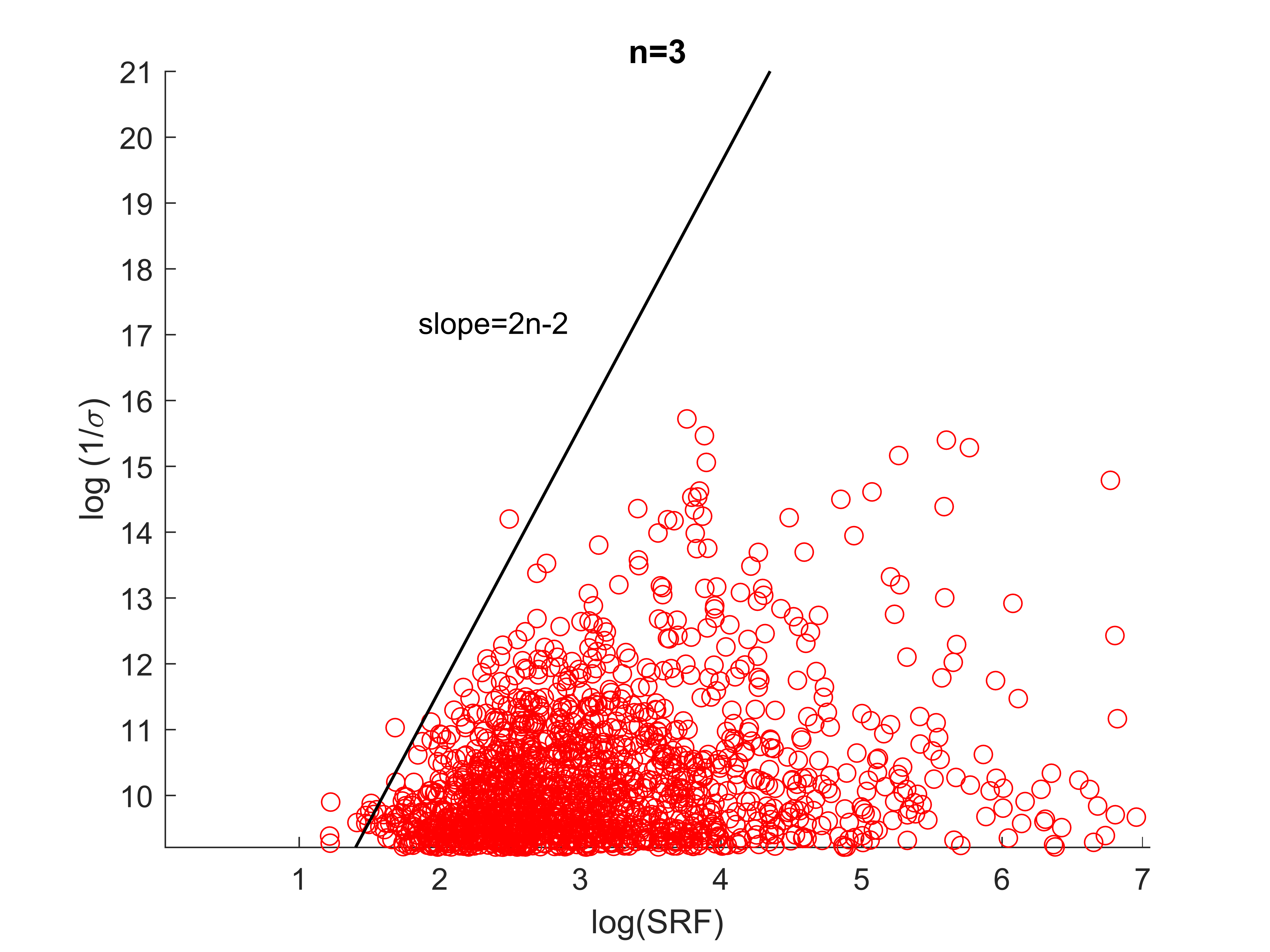}
		\caption{detection fail}
	\end{subfigure}
	\begin{subfigure}[b]{0.28\textwidth}
		\centering
		\includegraphics[width=\textwidth]{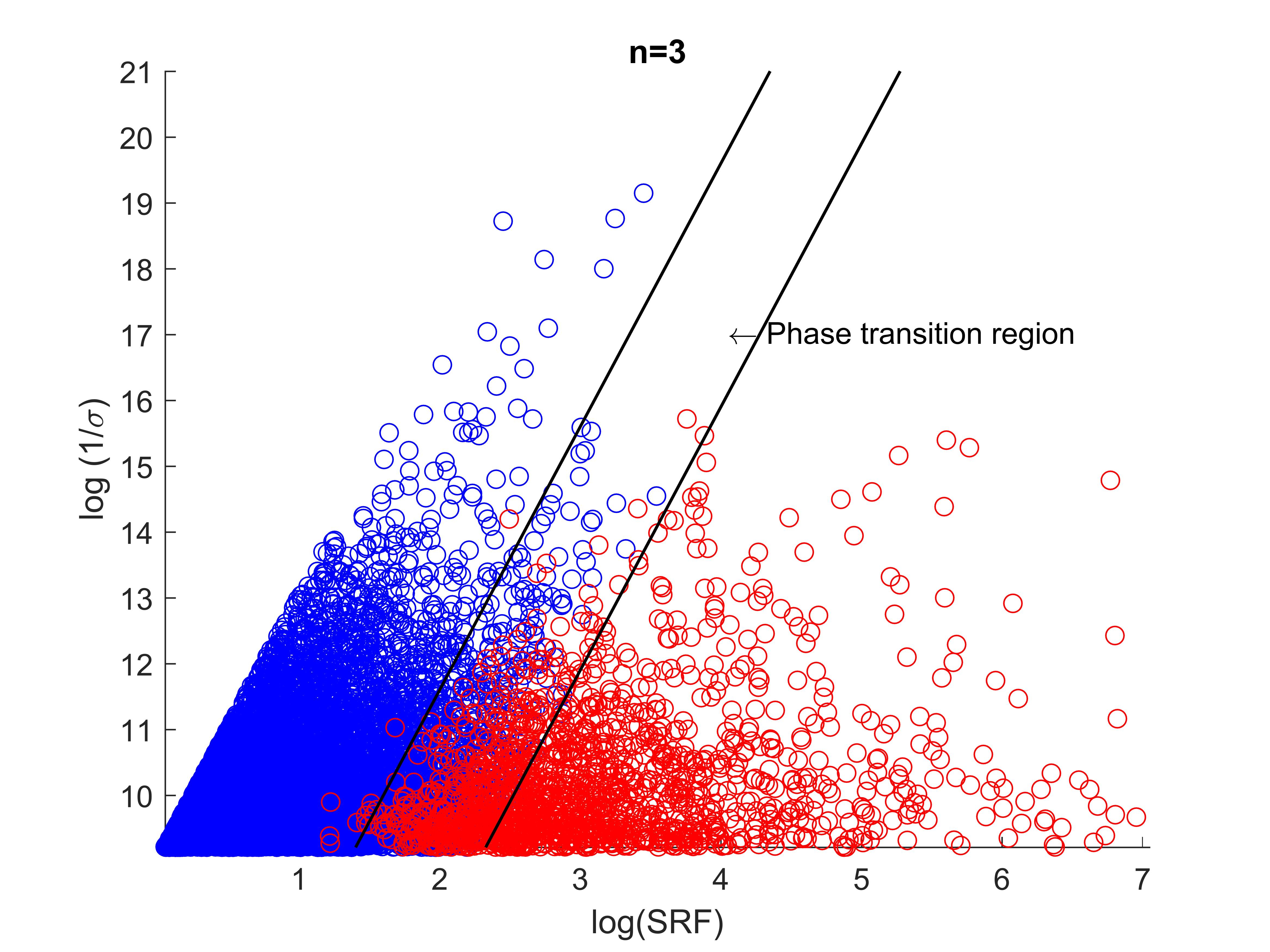}
		\caption{phase transition region}
	\end{subfigure}
	\begin{subfigure}[b]{0.28\textwidth}
		\centering
		\includegraphics[width=\textwidth]{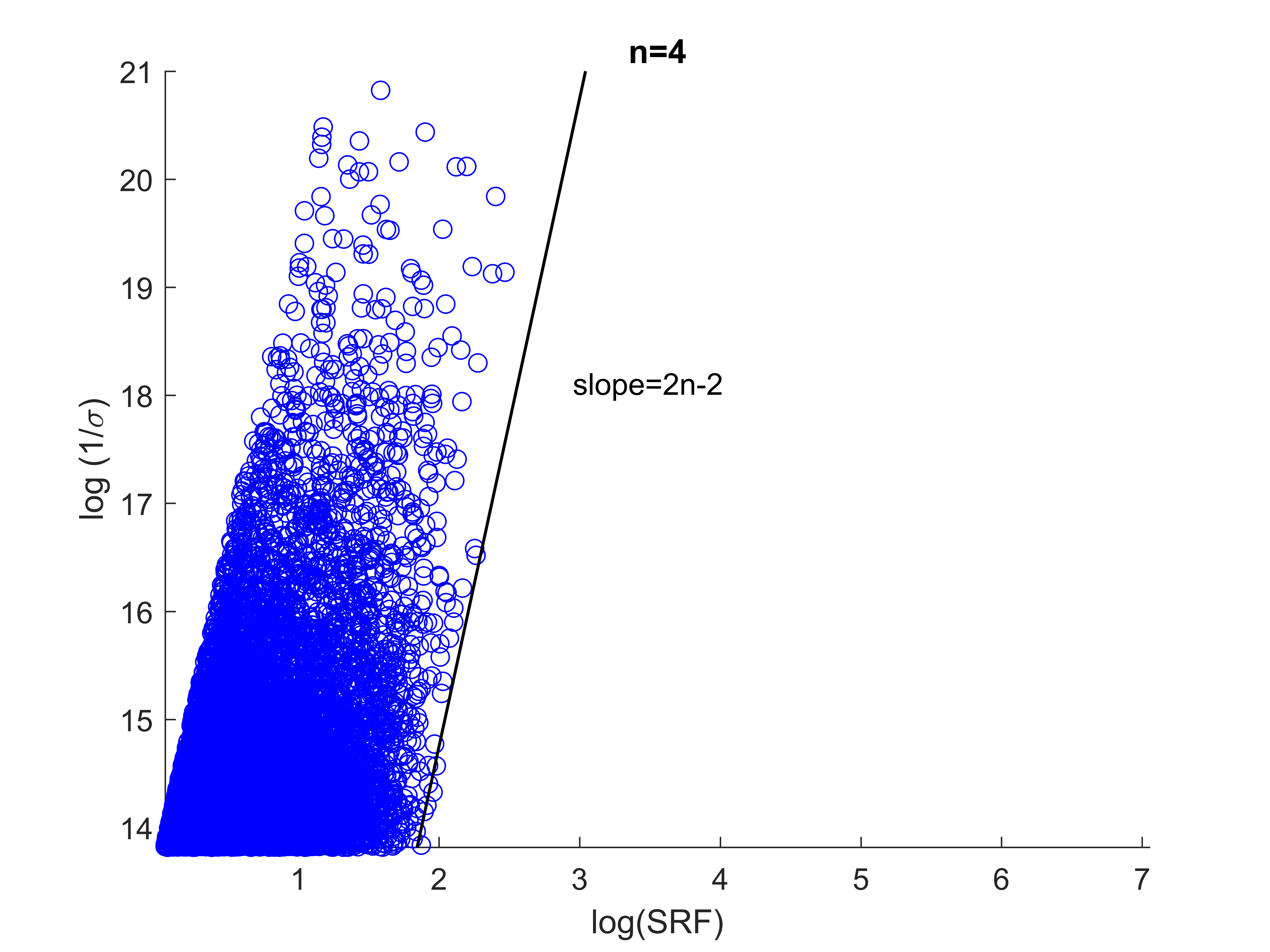}
		\caption{detection success}
	\end{subfigure}
	\begin{subfigure}[b]{0.28\textwidth}
		\centering
		\includegraphics[width=\textwidth]{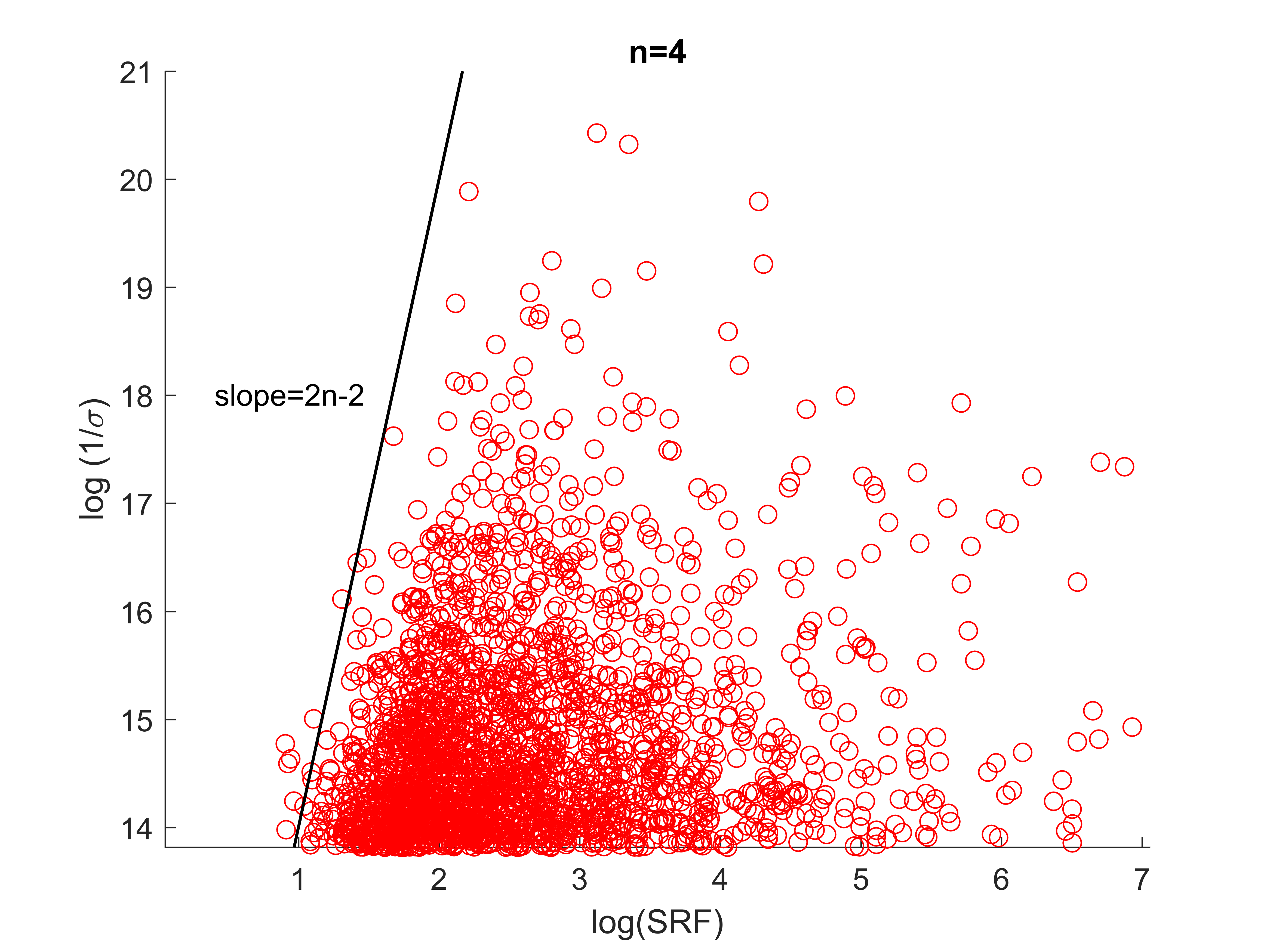}
		\caption{detection fail}
	\end{subfigure}
	\begin{subfigure}[b]{0.28\textwidth}
		\centering
		\includegraphics[width=\textwidth]{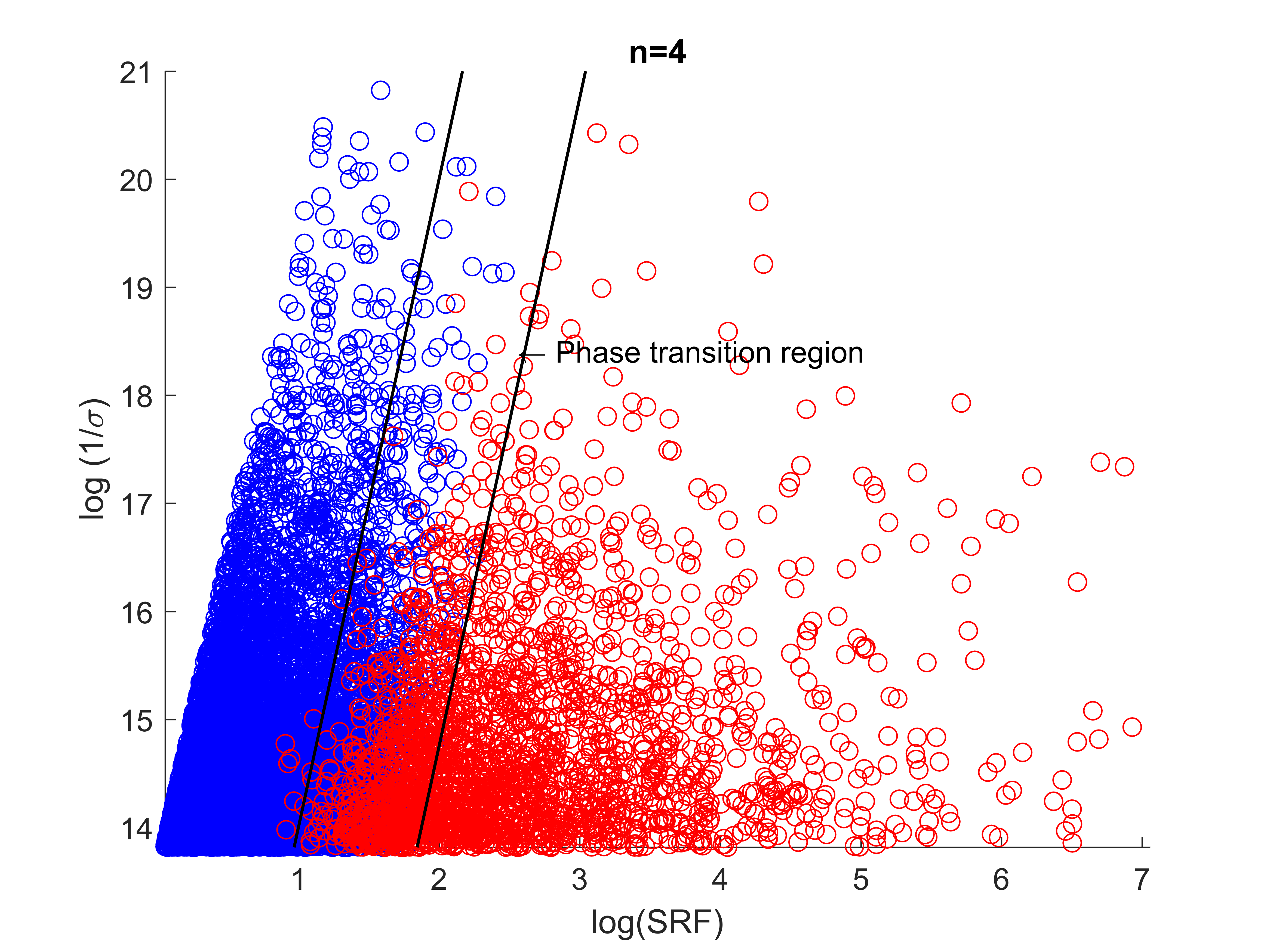}
		\caption{phase transition region}
	\end{subfigure}
	
	\caption{Plots of the successful and the unsuccessful number detection by \textbf{Algorithm \ref{algo:threedsweepingnumber}} depending on the relation between $\log(SRF)$ and $\log(\frac{1}{\sigma})$. (a) illustrates that four point sources can be exactly detected if $\log(\frac{1}{\sigma})$ is above a line of slope $4$ in the parameter space. Conversely, for the same case, (b) shows that the number detection fails if $\log(\frac{1}{\sigma})$ falls below another line of slope $4$. (f) highlights the phase transition region which is bounded by the black slashes in (a) and (b). (d),(e) and (f) illustrate parallel results for four point sources.}
	\label{fig:threednumberphasetransition}
\end{figure}

Figure \ref{fig:twodnumberphasetransition} and \ref{fig:threednumberphasetransition} shows the results of detecting source numbers in 2-dimensions and 3-dimensions respectively. It is shown that, in each case, two lines of slope $2n-2$ strictly separate the blue points (successful recoveries) and red points (unsuccessful recoveries) and in-between is the phase transition region. This phenomenon is exactly the one predicted by our theoretical analysis on the computational resolution limit of the number detection problem. It also manifests the efficiency of the two algorithms as they can resolve the source number correctly in the regime where the source separation distance is of the order of the computational resolution limit.

\section{Subspace projection based support recovery algorithms}\label{section:supportalgorithm}
In this section, we propose a subspace projection based support recovery algorithm in multi-dimensions. We remark that there are many algorithms that works efficiently to resolve point sources in one dimension. See for instance \cite{Prony-1795, schmidt1986multiple, stoica1989music, roy1989esprit, hua1990matrix, hua1991svd, denoyelle2017support, morgenshtern2016super, morgenshtern2020super}. Among them, subspace methods such as MUSIC, ESPRIT and Matrix Pencil method 
are shown to have favourable performance. A typical subspace method estimates the source locations based on the singular value decomposition of the data matrix, see for instance  \cite{li2019super, liao2016music}. 
In \cite{batenkov2019super}, the authors provide  numerical evidence that the Matrix Pencil method attains the performance bounds of the order of the computational resolution limit. 
In this paper, we shall use the Matrix Pencil method for the 1-dimensional problem. 


\subsection{Matrix Pencil Method for support recovery}\label{section:onedsupportalgorithm}

In this section, we review the Matrix Pencil method for 1-dimensional support recovery problem. Our presentation follows the one in  \cite{batenkov2019super}. 

Let $\vect H_u:= \vect H(s)[1:s, :]$ (and $H_l := H(s)[2:s+1, :]$) be the $s\times (s+1)$ matrix obtained from the Hankel matrix $\vect H(s)$ (\ref{equ:hankelmatrix1}) by selecting the first $s$ rows (respectively, the $2$ to $s+1$ rows). Then it turns out that, in the noiseless case,
$e^{i y_j \frac{\Omega}{s}}, 1\leq j \leq n$,  are exactly the nonzero generalized eigenvalues of the
pencil $\vect H_{l}-z\vect H_{u}$. In the noisy case, when the sources are well separated, each of the first $n$ nonzero generalized eigenvalues of the
pencil $\vect H_{l}-z\vect H_{u}$ is close to $e^{i y_j \frac{\Omega}{s}}$ for some $j$ \cite{moitra:2015:SEF:2746539.2746561}. We summarize the MP method in \textbf{Algorithm \ref{algo:onedmpsupport}} (see also \textbf{Algorithm 3.1} in \cite{batenkov2019super}), and the
interested reader is referred to the widely available literature on the subject (e.g. \cite{hua1991svd, hua1990matrix, moitra:2015:SEF:2746539.2746561},
and references therein) for further details. 

\begin{algorithm}[H]
\caption{\textbf{The Matrix Pencil algorithm}}\label{algo:onedmpsupport}	
\textbf{Input:} Source number $n$, measurement: $\mathbf{Y}(\omega), \vect \omega \in \mathbb R, ||\vect \omega||_2\leq \Omega$\\
1: $r=(M-1)\mod 2s$,  $\mathbf{Y}_{new}=(\mathbf Y(\omega_1), \mathbf Y(\omega_{r+1}), \cdots, \mathbf Y(\omega_{2sr+1}))^T$\;
2: Formulate the $(s+1)\times(s+1)$ Hankel matrix $\mathbf H(s)$ from $\mathbf{Y}_{new}$, and the matrices $\vect H_u, \vect H_l$\;
3: Compute the truncated Singular Value Decomposition (SVD) of $\vect H_u$, $\vect H_l$ of order $n$:
\[
\vect H_u=U_1\Sigma_1V_1^*,\quad \vect H_l = U_2\Sigma_2V_2^*,
\]
where $U_1, U_2, V_1, V_2$ are $s\times n$ and $\Sigma_1, \Sigma_2$ are $n \times n$\;
4: Generate the reduced pencil
\[
\mathbf {\hat H}_u =  U_2^* U_1\Sigma_1V_1^* V_2,\quad  \mathbf {\hat H}_l= \Sigma_2,
\]
where $\mathbf {\hat H}_u$, $\mathbf {\hat H}_l$ are $n\times n$\;
5: Compute the generalized eigenvalues $\{\hat z_j\}$ of the reduced pencil $(\mathbf {\hat H}_u, \mathbf {\hat H}_l)$, and put
$\{\hat y_j\}=\{\angle \hat z_j\}, j=1, \cdots, n$\;
\textbf{Return} $\{ \hat y_j\}$
\end{algorithm}

\subsection{Subspace projection based Matrix Pencil Method for support recovery}\label{section:highdsupportalgorithm}
In this section we propose a subspace projection based Matrix Pencil algorithm for the support recovery in multi-dimensions. We refer the readers to  (\cite{del1997matrix, swindlehurst1993azimuth, karthikeyan2015formulation, zoltowski1996closed, yilmazer2006matrix, wang2008tree, xi2014computationally, gu2015joint, wu2009doa, ye2009two})  for other algorithms in various settings in dimension two or three. 

As is indicated by the proof of Theorem \ref{thm:highdupperboundsupportlimit0}, when sources are well separated in $\mathbb R^{k+1}$, there exist two unit vectors $\vect v_1, \vect v_2$ so that the projection of source locations in the two $k$-dimensional subspaces $\vect v_1^\perp$ and $\vect v_2^\perp$ can be stably recovered simultaneously. We can then find the original source positions from their projections. To demonstrate the idea, We first consider the 
2-dimensional case. Let $N=\frac{(n+2)(n-1)}{2}$ and
\begin{equation}\label{equ:support2dvectors1}
\vect v_q=(\cos\phi_{1,q}, \ \sin {\phi_{1,q}})^T, \quad \phi_{1, q}= \frac{q\pi}{N}, 1\leq q\leq N.
\end{equation}
For each 1-dimensional subspace $\vect v_q^\perp$, we first recover the source number therein and choose only those where the recovered number is exactly $n$.
We then recover the projection of the source positions in each of those 1-dimensional subspace using the 1-dimensional Matrix Pencil method.
We choose two vectors, denoted by $\vect v_1, \vect v_2$, from those $\vect v_q$'s so that the recovered positions in $\vect v_1^\perp, \vect v_2^\perp$ have respectively the largest and second largest minimum separation distance. We remark that, when $N$ is large, one can require additionally that $\vect v_1$ is not too much correlated to $\vect v_2$, say $|\vect v_1\cdot \vect v_2 | \leq c$ for some constant $0<c<1$ to ensure that the reconstruction of 2-dimensional locations from their projections on $\vect v_1^\perp, \vect v_2^\perp$ is stable.


We next construct the original source locations from their projection on the 1-dimensional subspaces $\vect v_1^\perp$ and $\vect v_2^{\perp}$. This is usually called the pair matching in DOA problem that ad hoc schemes \cite{zoltowski1989sensor, johnson1991operational, chen1992direction, yilmazer2006matrix} are derived to associate the estimated azimuth and elevation angles. In our paper, this can be done in the following manner. From the projected locations on the subspaces $\vect v_1^\perp, \vect v_2^{\perp}$, we first form a grid of $n^2$ points $\vect z_{1,1}, \vect z_{1,2}, \cdots, \vect z_{n,n}$. It can be shown that the original source positions are close to these grid points. These grid points reduce the off-the-grid recovery problem to an on-the-grid one. 
We then employ an enumeration method to recover the source locations from these grid points. To be more specific, we define $G(\vect z_{1,j_1},\cdots, \vect z_{n,j_n}) = \big(e^{i \vect z_{1,j_1} \vect \omega} \  e^{i \vect z_{2,j_2} \vect \omega}\ \cdots \ e^{i \vect z_{n,j_n} \vect \omega}\big)$ and solve the following optimization problem by enumeration, 
 \begin{equation}\label{equ:supportrecoverminiequ1}
 \min_{\mathbf{\hat a},\pi \in \zeta(n)}||G(\vect z_{1,\pi_1},\cdots, \vect z_{n,\pi_n}) \mathbf{\hat a} - \vect Y ||_2,
 \end{equation} 
 where $\zeta(n)$ is the set of all permutations of $\{1, \cdots, n\}$. We note that the computational complexity of the enumeration is low when $n$ is not large. We summarize the algorithm in \textbf{Algorithm \ref{algo:twodmpsupport}} below. 
 

\begin{algorithm}[H]\label{algo:twodmpsupport}
\caption{\textbf{Two-dimensional subspace projection based support recovery algorithm}}	
\textbf{Input:} Noise level $\sigma$, source number $n$,and measurement: $\mathbf{Y}(\vect \omega), \vect \omega \in \mathbb R^2, ||\vect \omega||_2\leq \Omega$\\
\textbf{Input:} A large enough $N$, and corresponding $N$ unit vectors $\vect v(\phi), \phi \in \Big\{ \frac{\pi}{N}, \frac{2\pi}{N}, \cdots, \pi\Big\}$ \\
1:\For{$\phi \in \Big\{ \frac{\pi}{N}, \frac{2\pi}{N}, \cdots, \pi\Big\}$}{
	Input $\sigma$ and $\mathbf{Y}(\vect \omega), \vect \omega \in \vect v(\phi_1)^{\perp}$ to \textbf{Algorithm \ref{algo:onedsweepingnumber}} to recover the projected source number $\hat n$\;
	\If{$\hat n == n$}{
		Input $\sigma, \mathbf{Y}(\vect \omega), \vect \omega \in \vect v(\phi_1)^{\perp}$ and $\hat n$ to \textbf{Algorithm \ref{algo:onedmpsupport}}, save the output of \textbf{Algorithm \ref{algo:onedmpsupport}} as $\vect {\hat p}_{1}, \cdots, \mathbf {\hat p}_{n}$\;} 
}
2: Choose two vectors, denoted by $\vect v_1, \vect v_2$, from those $\vect v(\phi)$'s so that the recovered positions $\vect {\hat y}_{j}(\vect v_1^\perp)$'s, $\mathbf {\hat y}_{j}(\vect v_2^\perp)$'s in spaces $\vect v_1^\perp, \vect v_2^\perp$ have respectively the largest and second largest minimum separation distance\;

3:Construct the $n^2$ grid points $\vect z_{1,1}, \vect z_{1,2}, \cdots, \vect z_{n,n}$ by considering the intersection points of lines $\mathbf {\hat y}_{q}(\vect v_j^\perp)+ \lambda \vect v_j, \ \lambda \in \mathbb R, \ q =1,\cdots,n, \ j =1,2$\;
4: Solve the following optimization problem by enumeration, 
\[\min_{\mathbf{\hat a},\pi \in \zeta(n)}||G(\vect z_{1,\pi_1},\cdots, \vect z_{n,\pi_n})\mathbf{\hat a} - \vect Y ||_2
\]
where $\zeta(n)$ is the set of all permutations of $\{1,\cdots, n\}$ and $G(\vect z_{1,\pi_1},\cdots, \vect z_{n,\pi_n}) = \big(e^{i \vect z_{1,\pi_1} \vect \omega} \ e^{i \vect z_{2,\pi_2} \vect \omega} \ \cdots \ e^{i \vect z_{n,\pi_n} \vect \omega}\big)$\;

6: The minimizer $\vect z_{1,\pi_1}, \cdots, \vect z_{n, \pi_n}$'s are the recovered source locations $\mathbf{\hat y}_1, \cdots, \mathbf{\hat y}_n$\;
\textbf{Return} $\mathbf{\hat y}_1, \cdots, \mathbf{\hat y}_n$.
\end{algorithm}

\medskip
Now we present the algorithm for 3-dimensional support recovery. 
For point sources located in dimension three, we first recover their projections in two properly chosen 2-dimensional subspaces. Precisely, we choose the unit vectors
\begin{equation}
\vect v(\phi_1, \phi_{2})=(\cos\phi_1, \ \sin {\phi_1}\cos {\phi_2},\ \sin \phi_1 \sin \phi_2)^T, \quad \phi_1, \phi_2\in \Big\{ \frac{\pi}{2N}, \frac{2\pi}{2N}, \cdots, \frac{\pi}{2}\Big\},
\end{equation}
where $N= \lfloor (\frac{n(n-1)}{2})^{\frac{1}{2}}\rfloor+1$. Utilizing measurement in each of the 2-dimensional subspaces $\vect v(\phi_1, \phi_{2})^{\perp}$, we can recover the source locations in some of these spaces stably by \textbf{Algorithm \ref{algo:twodmpsupport}}. We choose the two subspaces where the minimum separation distance of the reconstructed project source locations is the largest and second largest and denote them by $\vect v_1^\perp$ and $\vect v_2^\perp$ respectively. 
The recovered locations in these two subspaces are denoted by $\mathbf {\hat y}_{j}(\vect v_1^\perp)$, $1\leq j \leq n$,  and $\mathbf {\hat y}_{j}(\vect v_2^\perp)$, $1\leq j \leq n$, respectively. For each pair of $j, p$, we construct two lines $\vect {\hat y}_{j}(\vect v_1^\perp)+ \lambda \vect v_1$ and $\vect {\hat y}_{p}(\vect v_2^\perp)+ \lambda \vect v_2$ and denote them by $l_{1,j}, l_{2,p}$ respectively. For the line pair $(l_{1,j}, l_{2,p})$, we find the two  points $\vect c_{1, j} \in l_{1,j}, \vect c_{2, p} \in l_{2,j}$ which minimize the distance of the two lines.  We then choose the middle point $\frac{\vect c_{1,j}+\vect c_{2,p}}{2}$ as a possible source location if $\|\vect c_{1, j}- \vect c_{2, p}\|_2$ is not great than the two minimum separation distances in $\vect v_1^\perp$ and $\vect v_2^\perp$. Finally by solving an optimization problem with these possible source locations, we can reconstruct stably the original source positions. The procedure is similar to the two dimensional case and is summarized in \textbf{Algorithm \ref{algo:threedmpsupport}} below. 


\begin{algorithm}[H]\label{algo:threedmpsupport}
	\caption{\textbf{Three-dimensional subspace projection based support recovery algorithm}}	
	\textbf{Input:} Noise level $\sigma$, source number $n$,and measurement: $\mathbf{Y}(\vect \omega), \vect \omega \in \mathbb R^2, ||\vect \omega||_2\leq \Omega$\\
	\textbf{Input:} A large enough $N$, and corresponding $N^2$ unit vectors $\vect v(\phi_1, \phi_2), \phi_1, \phi_2 \in \Big\{ \frac{\pi}{2N}, \frac{2\pi}{2N}, \cdots, \frac{\pi}{2}\Big\}$ \\
	1:\For{$\phi_1, \phi_2 \in \Big\{ \frac{\pi}{2N}, \frac{2\pi}{2N}, \cdots, \frac{\pi}{2}\Big\}$}{
		Input $\sigma$ and $\mathbf{Y}(\vect \omega), \vect \omega \in \vect v(\phi_1, \phi_2)^{\perp}$ to \textbf{Algorithm \ref{algo:twodsweepingnumber}} to recover the projected source number $\hat n$\;
		\If{$\hat n == n$}{
		Input $\sigma, \mathbf{Y}(\vect \omega), \vect \omega \in \vect v(\phi_1, \phi_2)^{\perp}$ and $\hat n$ to \textbf{Algorithm \ref{algo:twodmpsupport}},
		save the output as $\vect {\hat p}_{1}, \cdots, \mathbf {\hat p}_{n}$\;} 
	}
	2:Choose two vectors, denoted by $\vect v_1, \vect v_2$, from those $\vect v(\phi_1, \phi_2)$'s so that the recovered projected positions $\vect {\hat y}_{j}(\vect v_1^\perp)'s, \mathbf {\hat y}_{j}(\vect v_2^\perp)'s$ in spaces $\vect v_1^{\perp}, \vect v_2^{\perp}$ have respectively the largest and the second largest minimum separation distance (denoted by $d_{\min}, \hat d_{\min}$ respectively)\;
	3: Denote the line $\vect {\hat y}_{j}(\vect v_1^\perp)+ \lambda \vect v_1$ by $l_{1,j}$ and the line $\vect {\hat y}_{p}(\vect v_2^\perp)+ \lambda \vect v_2$ by $l_{2,p}$. For every pair of lines $l_{1,j}, l_{2,p}$, find the two nearest points in the two lines and denote them by $\vect c_{1, j}, \vect c_{2, p}$. If $||\vect c_{1,j}-\vect c_{2,p}||< \min(d_{\min}, \hat d_{\min})$, consider the point $\frac{\vect c_{1,j}+\vect c_{2,p}}{2}$ as a candidate location and denote it by $\vect z_{j,p}$. Ignore those $\frac{\vect c_{1,j}+\vect c_{2,p}}{2}$ if $||\vect c_{1,j}-\vect c_{2,p}||\geq \min(d_{\min}, \hat d_{\min})$\; 
	4: Solve the following optimization problem over the above $\vect z_{j,p}$'s, 
\[\min_{\mathbf{\hat a}, j_p \neq j_q, p\neq q}||G(\vect z_{1,j_1},\cdots, \vect z_{n,j_n})\mathbf{\hat a} - \vect Y ||_2
\]
where $G(\vect z_{1,j_1},\cdots, \vect z_{n,j_n}) = \big(e^{i \vect z_{1,j_1} \vect \omega} \  e^{i \vect z_{2,j_2} \vect \omega}\ \cdots \ e^{i \vect z_{n,j_n} \vect \omega}\big)$\;
5: The minimizer $\vect z_{1,j_1}, \cdots, \vect z_{n, j_n}$'s are the recovered source locations $\mathbf{\hat y}_1, \cdots, \mathbf{\hat y}_n$\;
\textbf{Return} $\mathbf{\hat y}_1, \cdots, \mathbf{\hat y}_n$.
\end{algorithm}

\subsection{Numerical experiments and phase transition}
\begin{figure}[!h]
	\centering
	\begin{subfigure}[b]{0.28\textwidth}
		\centering
		\includegraphics[width=\textwidth]{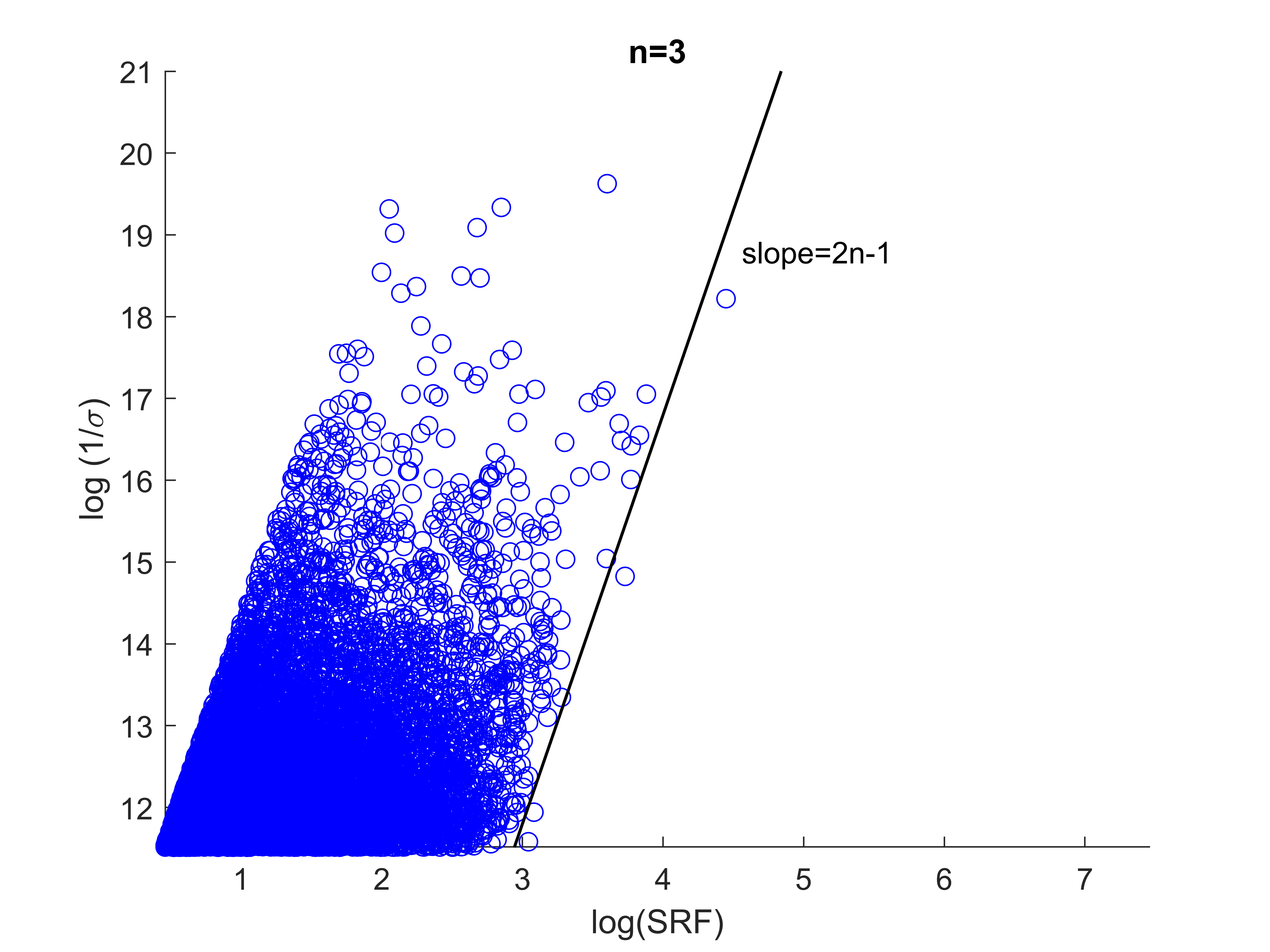}
		\caption{recovery success}
	\end{subfigure}
	\begin{subfigure}[b]{0.28\textwidth}
		\centering
		\includegraphics[width=\textwidth]{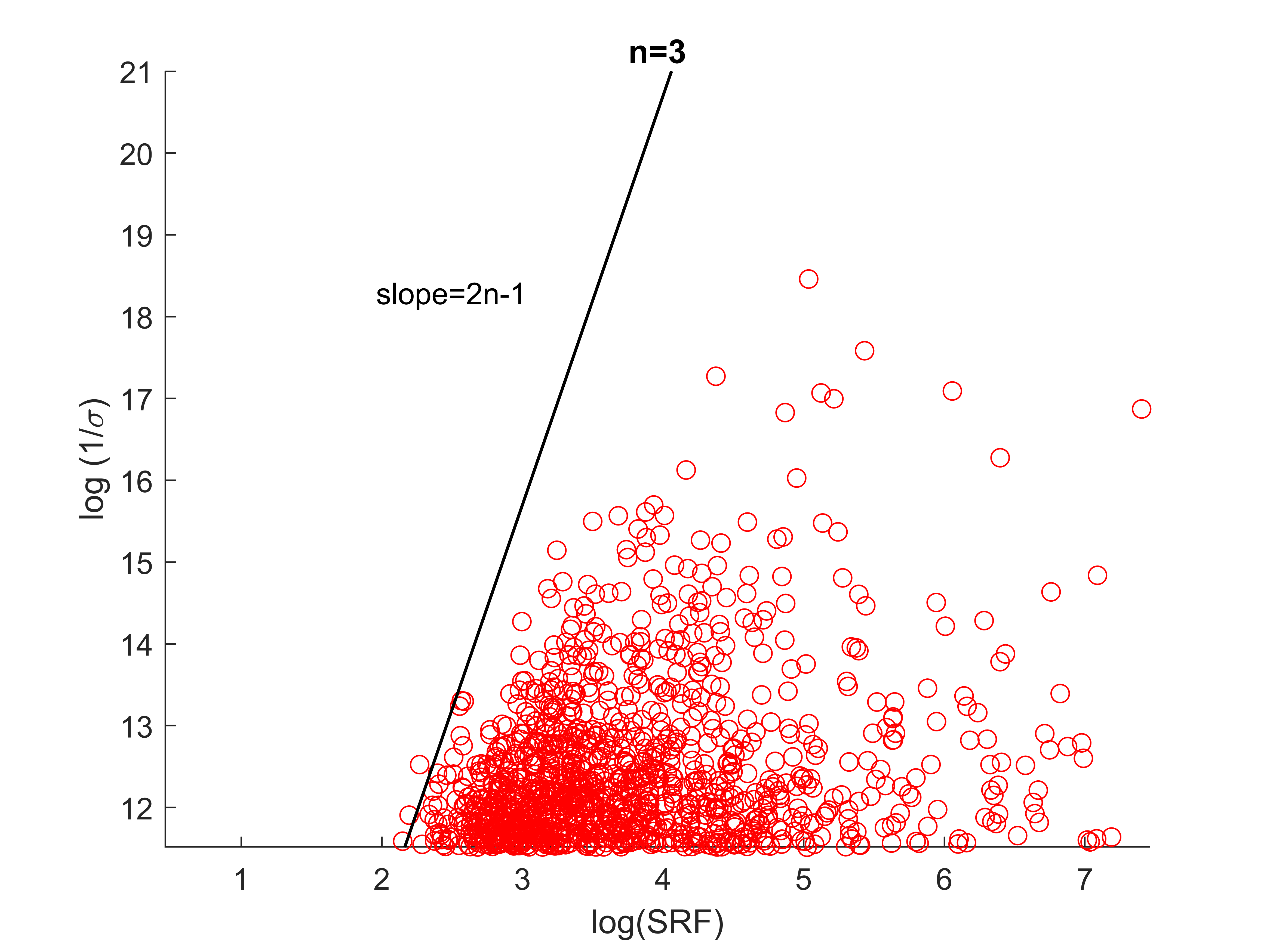}
		\caption{recovery fail}
	\end{subfigure}
	\begin{subfigure}[b]{0.28\textwidth}
		\centering
		\includegraphics[width=\textwidth]{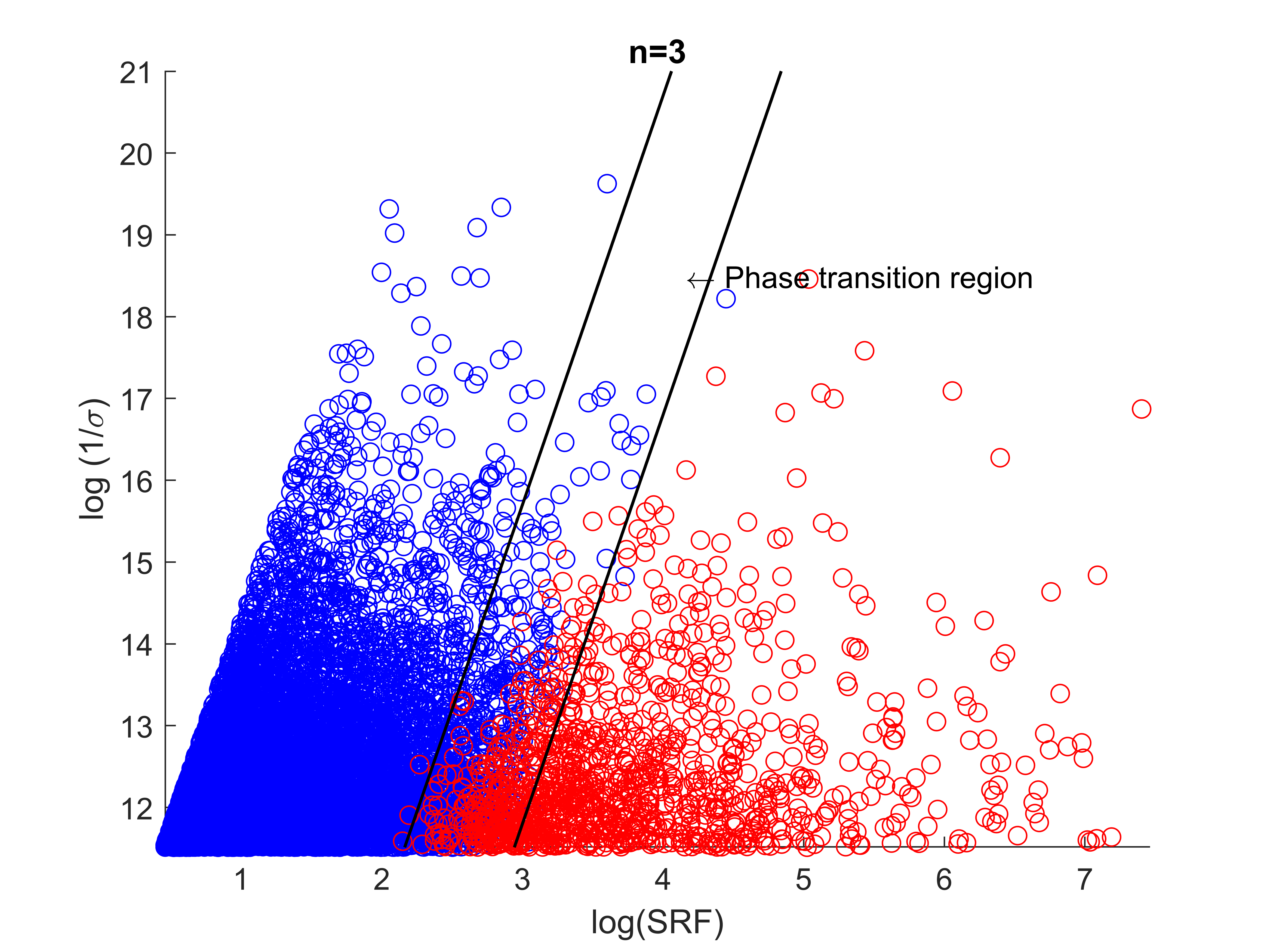}
		\caption{phase transition region}
	\end{subfigure}
	\begin{subfigure}[b]{0.28\textwidth}
		\centering
		\includegraphics[width=\textwidth]{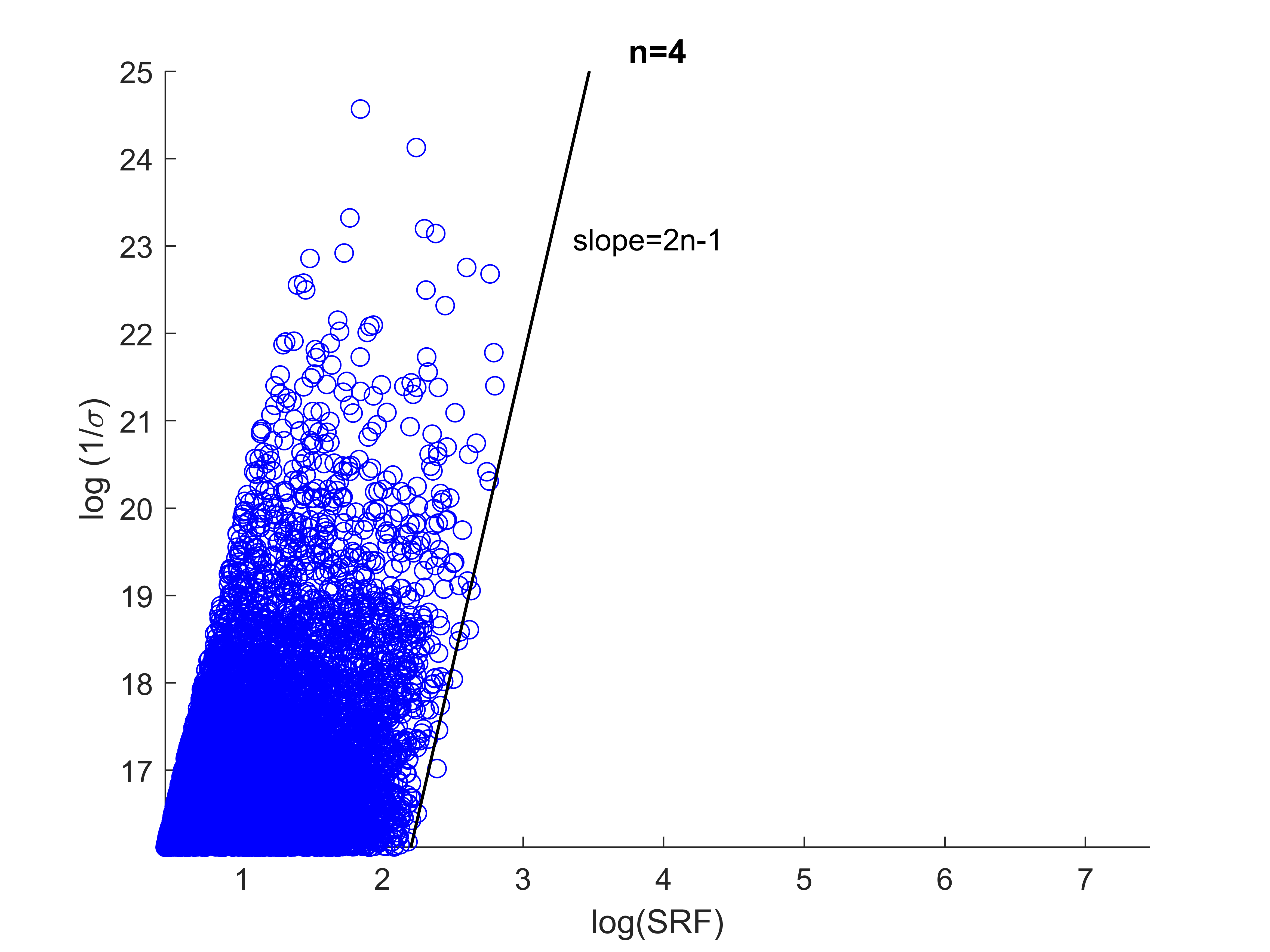}
		\caption{recovery success}
	\end{subfigure}
	\begin{subfigure}[b]{0.28\textwidth}
		\centering
		\includegraphics[width=\textwidth]{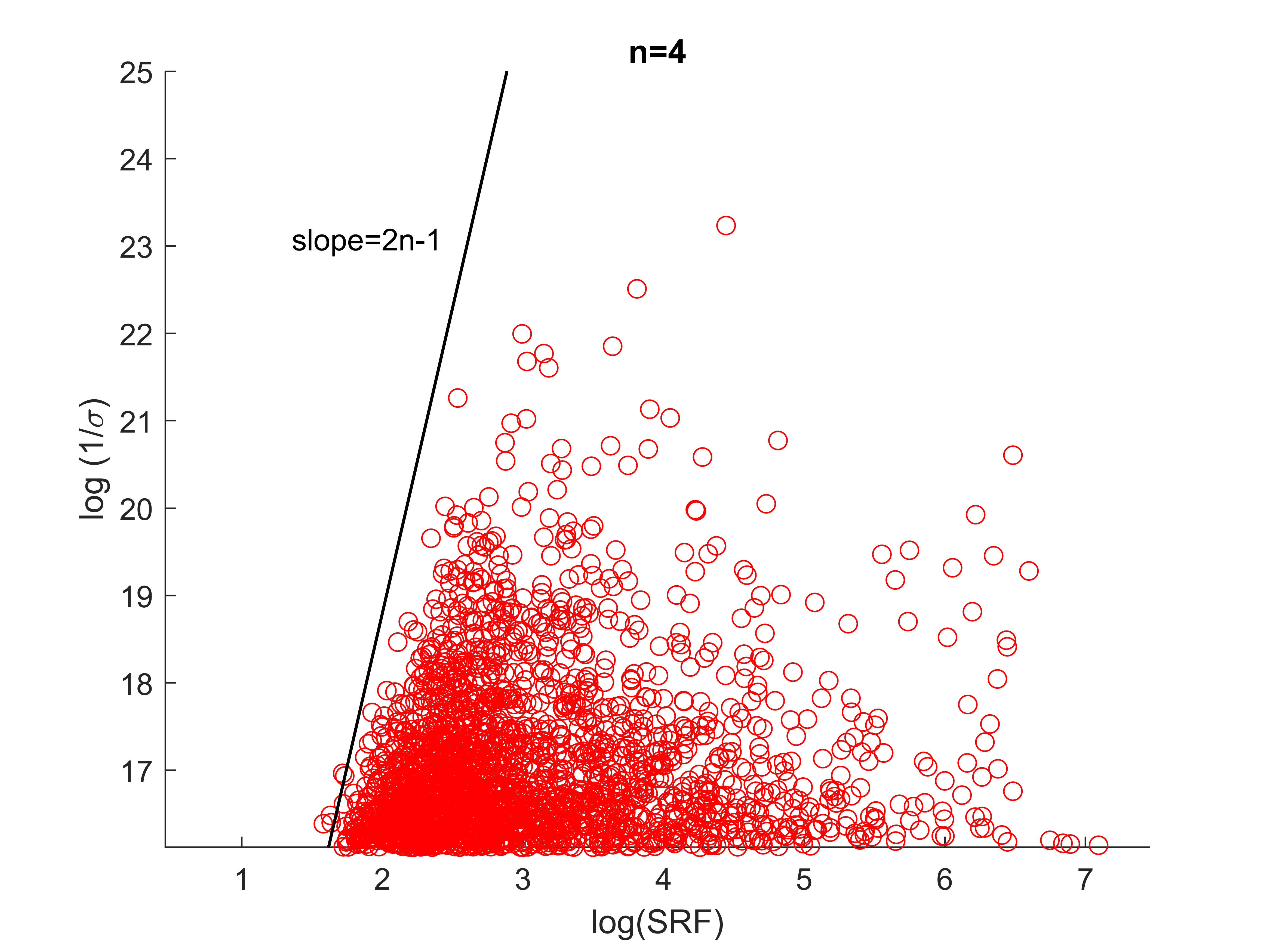}
		\caption{recovery fail}
	\end{subfigure}
	\begin{subfigure}[b]{0.28\textwidth}
		\centering
		\includegraphics[width=\textwidth]{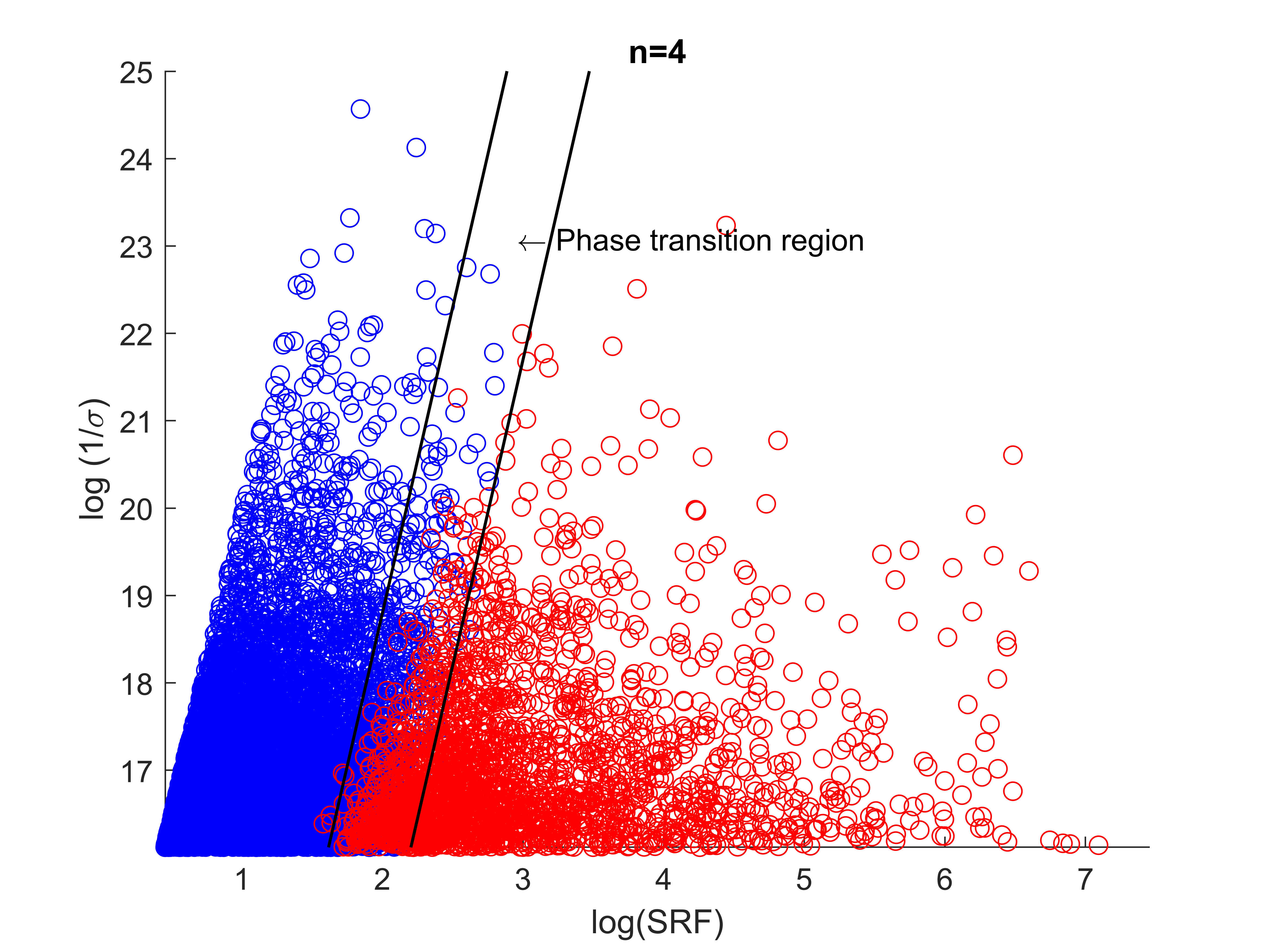}
		\caption{phase transition region}
	\end{subfigure}
	\caption{Plots of the successful and the unsuccessful support recovery by \textbf{Algorithm \ref{algo:twodmpsupport}} depending on the relation between $\log(SRF)$ and $\log(\frac{1}{\sigma})$. (a) illustrates that locations of three point sources can be stably recovered if $\log(\frac{1}{\sigma})$ is above a line of slope $5$ in the parameter space. Conversely, for the same case, (b) shows that the 2-dimensional support recovery fails if $\log(\frac{1}{\sigma})$ falls below another line of slope $5$. (f) highlights the phase transition region which is bounded by the black slashes in (a) and (b). (d),(e) and (f) illustrate parallel results for four point sources.}
	\label{fig:twodsupportphasetransition}
\end{figure}

\begin{figure}[!h]
	\centering
	\begin{subfigure}[b]{0.28\textwidth}
		\centering
		\includegraphics[width=\textwidth]{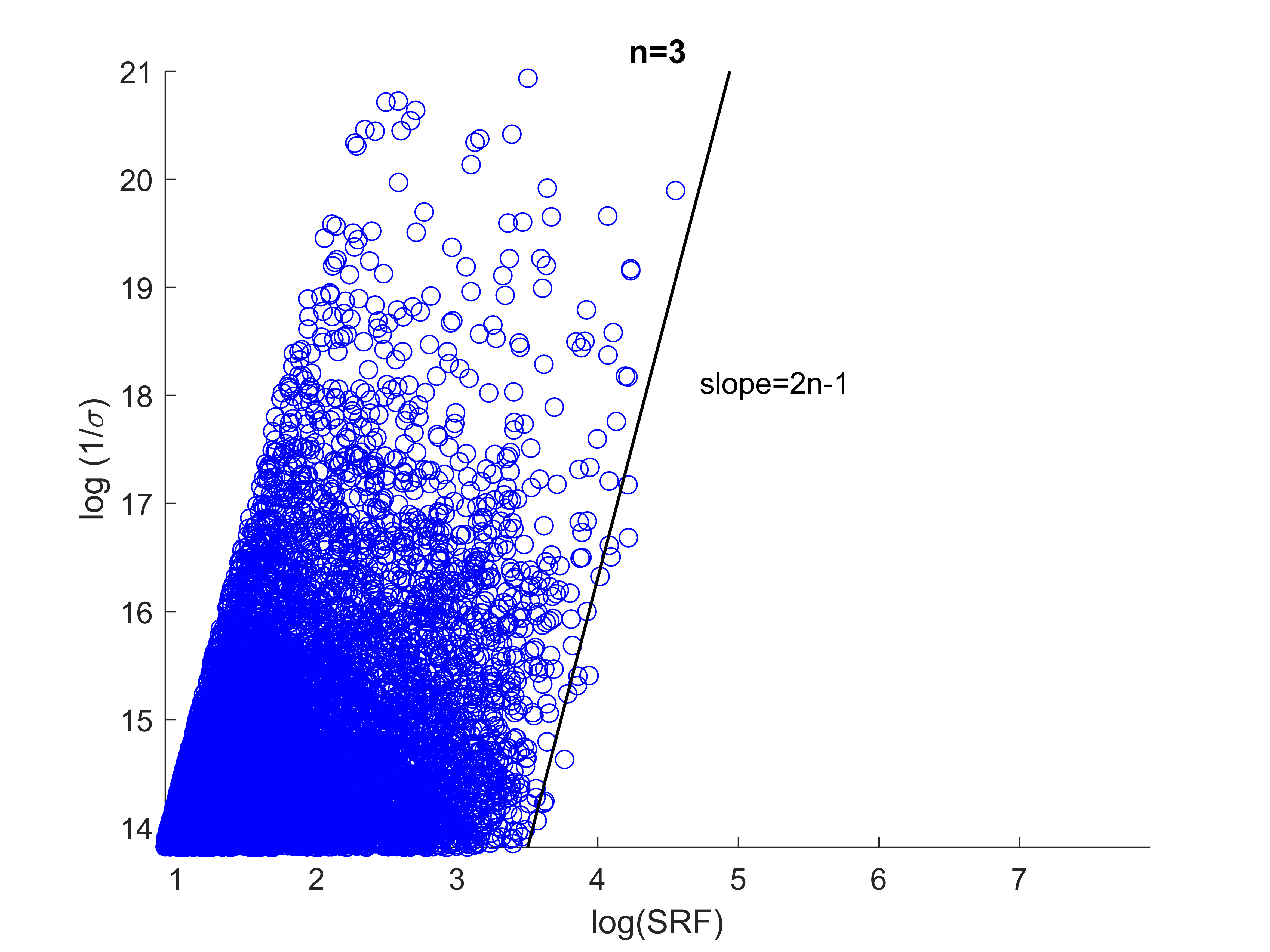}
		\caption{recovery success}
	\end{subfigure}
	\begin{subfigure}[b]{0.28\textwidth}
		\centering
		\includegraphics[width=\textwidth]{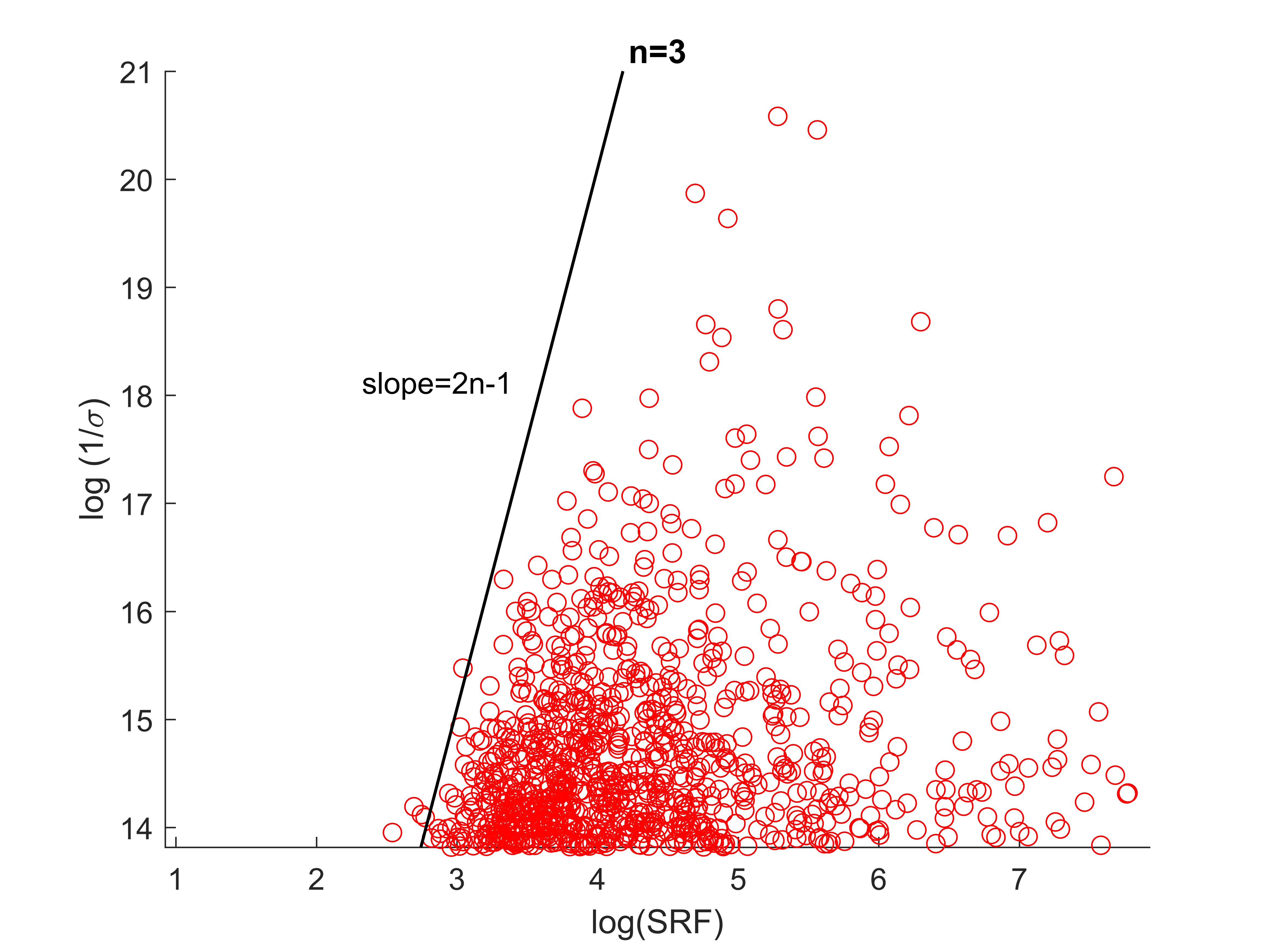}
		\caption{recovery fail}
	\end{subfigure}
	\begin{subfigure}[b]{0.28\textwidth}
		\centering
		\includegraphics[width=\textwidth]{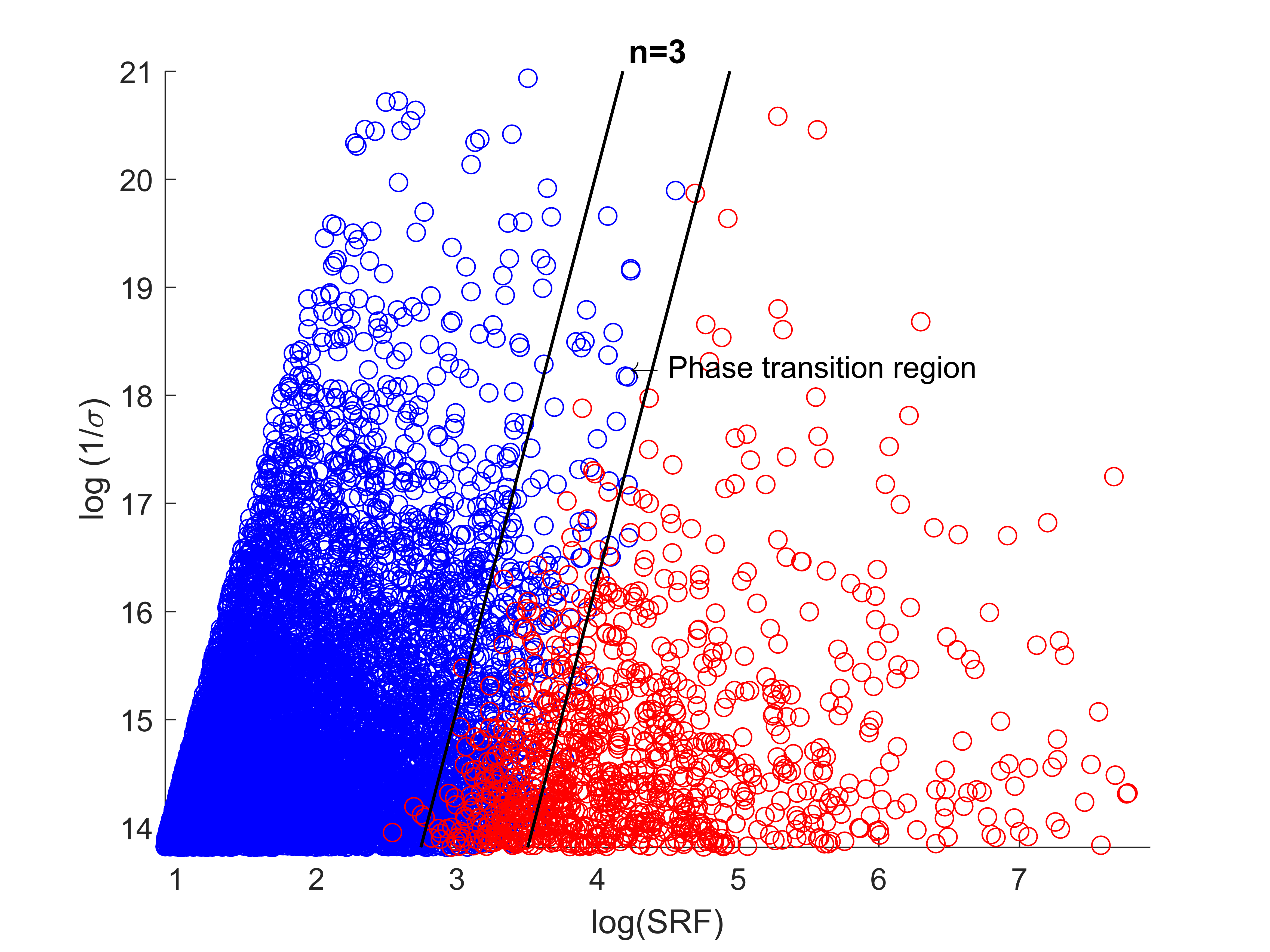}
		\caption{phase transition region}
	\end{subfigure}
	\begin{subfigure}[b]{0.28\textwidth}
		\centering
		\includegraphics[width=\textwidth]{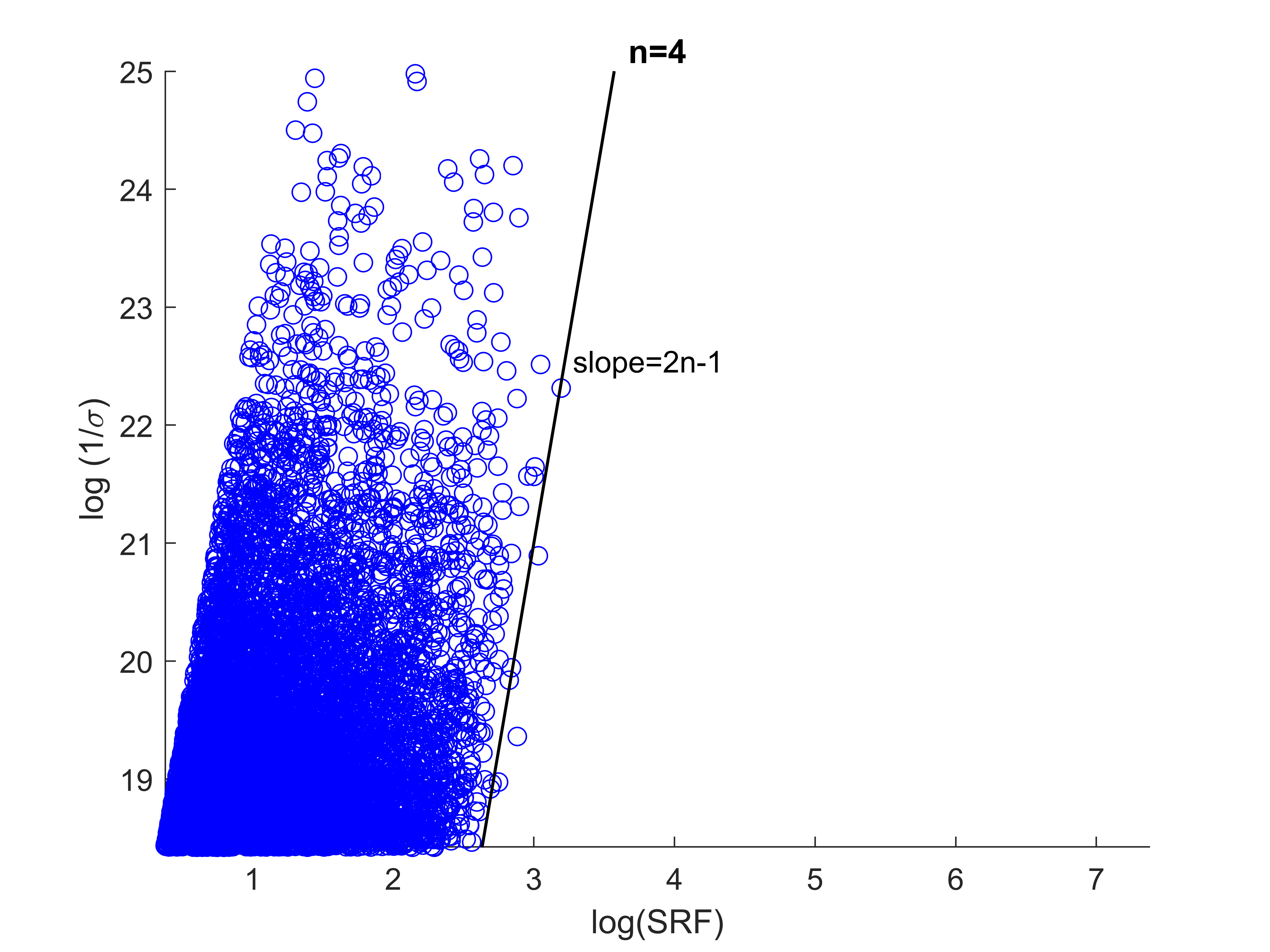}
		\caption{recovery success}
	\end{subfigure}
	\begin{subfigure}[b]{0.28\textwidth}
		\centering
		\includegraphics[width=\textwidth]{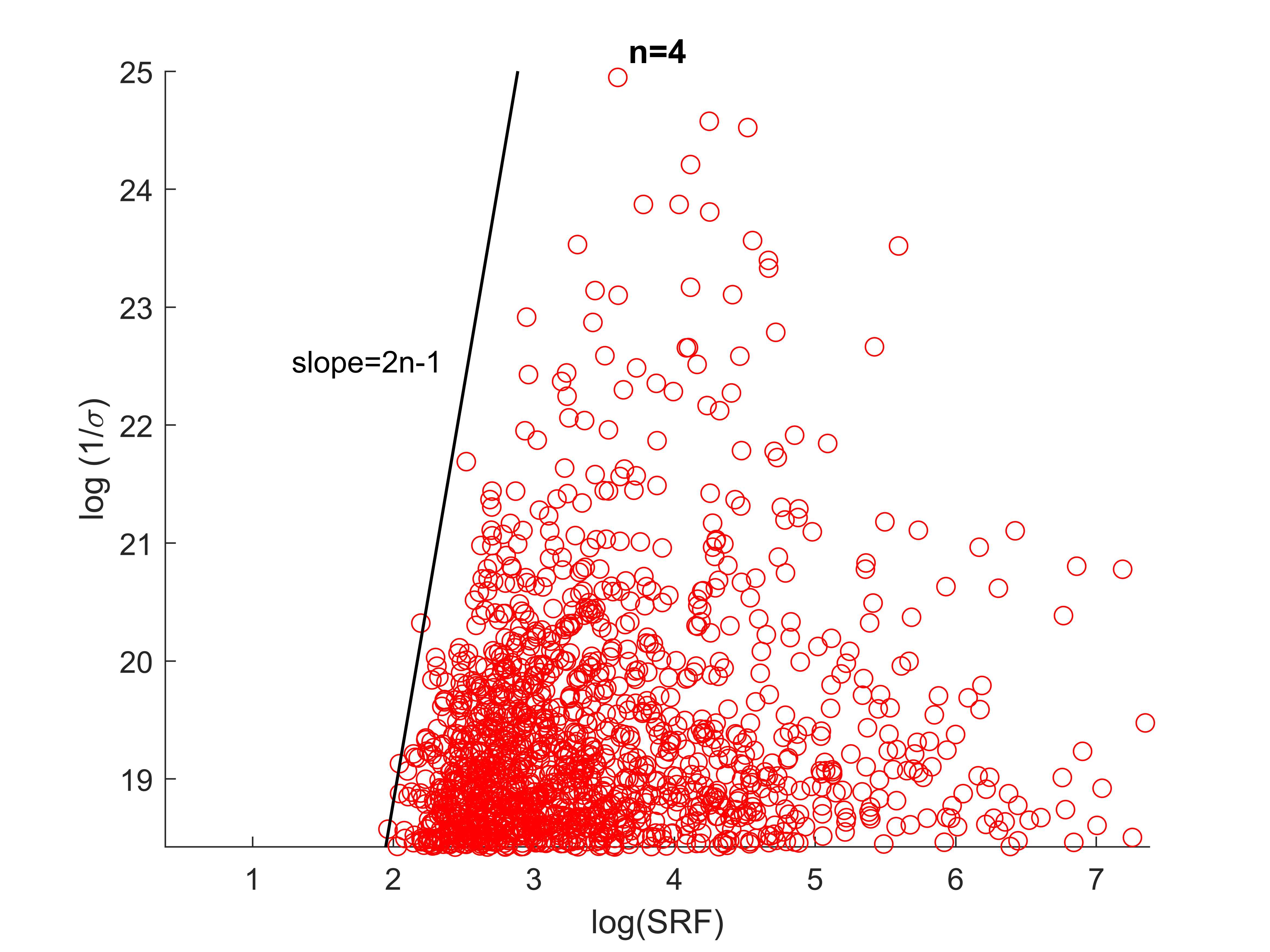}
		\caption{recovery fail}
	\end{subfigure}
	\begin{subfigure}[b]{0.28\textwidth}
		\centering
		\includegraphics[width=\textwidth]{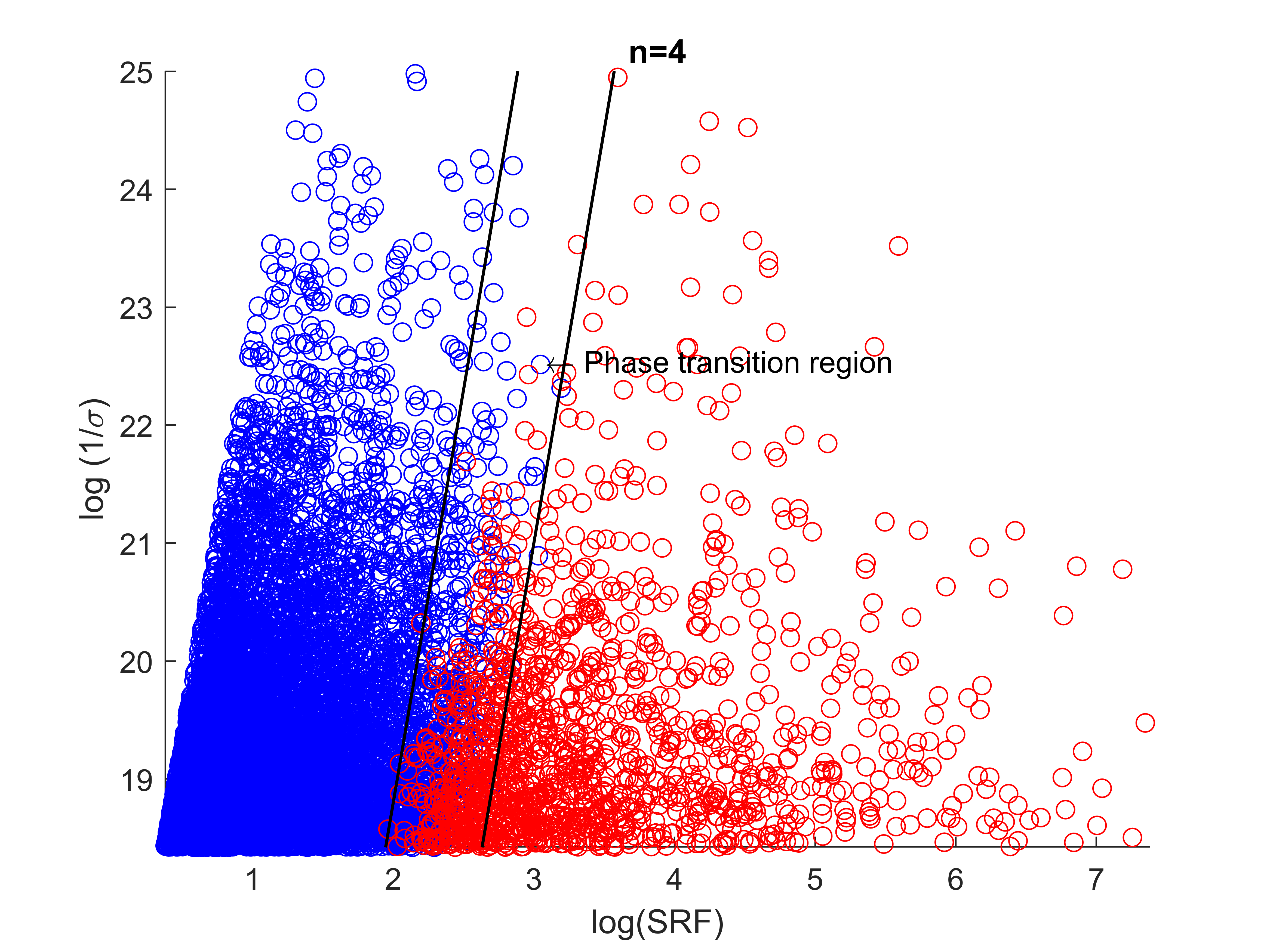}
		\caption{phase transition region}
	\end{subfigure}
	\caption{Plots of the successful and the unsuccessful support recovery by \textbf{Algorithm \ref{algo:threedmpsupport}} depending on the relation between $\log(SRF)$ and $\log(\frac{1}{\sigma})$. (a) illustrates that locations of three point sources can be stably recovered if $\log(\frac{1}{\sigma})$ is above a line of slope $5$ in the parameter space. Conversely, for the same case, (b) shows that the 3-dimensional support recovery fails if $\log(\frac{1}{\sigma})$ falls below another line of slope $5$. (f) highlights the phase transition region which is bounded by the black slashes in (a) and (b). (d),(e) and (f) illustrate parallel results for four point sources.}
	\label{fig:threedsupportphasetransition}
\end{figure}

We perform numerical experiments to demonstrate the phase transition phenomenon regarding to the super-resolution factor and the SNR for the support recovery in dimension two and three. 
In view of the computational resolution limit for support recovery (which is of the order $O(\frac{1}{\Omega}\big(\frac{\sigma}{m_{\min}}\big)^{\frac{1}{2n-1}})$), recovering $5$ closely-spaced point sources demands extremely low noise level. Therefore, in the experiments we only consider the case of three and four sources.
For the $2$-dimensional case, we fix $\Omega=1$ and  consider $n$ ($n=3$ or $4$) point sources separated with minimum separation $d_{\min}$. We perform 10000 random experiments (the randomness is in the choice of $(d_{\min},\sigma, \vect y_j, a_j)$ to recover the source locations using \textbf{Algorithm \ref{algo:singleexperiemnt}}. For the $3$-dimensional case, we conduct $10000$ random experiments under the same setup. 
As is shown in Figure \ref{fig:twodsupportphasetransition} and \ref{fig:threedsupportphasetransition}, in each case, two lines with slope $2n-1$ strictly separate the blue points (successful cases) and red points (unsuccessful cases), and in-between is the phase transition region. This is exactly the predicted phase transition phenomenon by our theory of computational resolution limit for support recovery. It also demonstrates that the proposed support recovery algorithm can resolve the location of point sources in the regime where the separation distance is on the order of the computational resolution limit.

\begin{algorithm*}[H]\label{algo:singleexperiemnt}
	\caption{\textbf{A single experiment}}	
	\textbf{Input:} Sources $\mu=\sum_{j=1}^{n}a_j \delta_{\vect y_j}$, Noise level $\sigma$\\
	\textbf{Input:} Measurements: $\mathbf{Y}(\vect \omega), ||\vect \omega||_2\leq \Omega$\\
	1: $\text{Successnumber}=0$\;
	2: Input source number $n$ and measurement $\vect Y$ to \textbf{Algorithm \ref{algo:twodmpsupport}} (\textbf{Algorithm \ref{algo:threedmpsupport}} for $3$-dimensional case) and save the output as $\vect y_1, \cdots, \vect y_n$\; 
	\For{each $1\leq j \leq n$}{
		Compute the error for the source location $\vect y_j$:
		$e_j:=\min_{\mathbf{\hat y}_l, l=1, \cdots,n }||\mathbf{\hat y}_l- \vect y_j||_2$\; 
		The source location $\vect y_j$ is recovered successfully if
		\[e_j< \frac{\min_{p\neq j}||\vect y_p- \vect y_j||_2}{3};\]
		and 
		\[\text{Successnumber}=\text{Successnumber}+1;\] 
	}  
	\eIf{$\text{Successnumber}==n$}{
		Return Success}
	{Return Fail}
\end{algorithm*}

\newpage
\section{Appendix}\label{section:appendix}
In this section, we present and prove some technical lemmas that are used in the subspace projection strategy we used to prove the main results on the computational resolution limit in multi-dimensions. 
We denote the unit sphere in $\mathbb R^k$ as $S_{k-1}$ and its area as $\text{area}(S_{k-1})$. For each $\vect u \in \mathbb R^k$, we denote

\begin{align}\label{equ:defineofN}
N(\vect{u}, \Delta) &= \Big\{ \vect v \Big|\vect v \in S_{k-1}, ||\mathcal P_{\vect v^\perp}(\vect{u})||_2 <||\vect{u}||_2 \sin\Delta \Big\}. 
\end{align}
It can be verified that
\[
N(\vect{u}, \Delta)= \Big\{ \vect v \ \Big| \vect v \in S_{k-1}, |\vect{u}\cdot \vect v| > ||\vect{u}||_2 \cos
\Delta \Big\}.
\]

\begin{lem}\label{lem:badareaestiamte}
For $0<\Delta \leq \frac{\pi}{2}$, we have 
$
\text{area}(N(\vect{u}, \Delta)) \leq \frac{2\Delta^{k-1}}{k-1} \text{area}(S_{k-2}).
$
\end{lem}
Proof: WLOG, we consider the case when the unit vector $\vect u= (1, 0, \cdots,0)^T$.  Using spherical coordinates (\ref{equ:sphericalcoordinates1}), the points in $S_{k-1}$ so that $|\vect u\cdot \vect v|> \cos \Delta$ can be expressed by
\[
\vect v = (\cos \phi_1,\  \sin \phi_1 \cos \phi_2,\  \cdots,\  \sin \phi_1 \cdots \sin \phi_{k-2}\cos\phi_{k-1},\  \sin \phi_1 \cdots \sin \phi_{k-2}\sin\phi_{k-1})^T
\]
where $0\leq \phi_1 < \Delta$ or $\pi -\Delta< \phi_1 \leq \pi$, and $\phi_2,\cdots, \phi_{k-2} \in [0, \pi], \phi_{k-1}\in [0, 2\pi]$. Therefore,  
\begin{align*}
\text{area}(N(\vect{u}, \Delta))=	&2\int_{0}^{2\pi}\int_{0}^{\pi}\cdots\int_{0}^{\Delta} \sin^{k-2}(\phi_1)\sin^{k-3}(\phi_2)\cdots \sin(\phi_{k-1})  d\phi_1 d\phi_2 \cdots d\phi_{k-2} d\phi_{k-1} \\
	= &2\text{area}(S_{k-2}) \int_{0}^{\Delta} \sin^{k-2}(\phi_1) d\phi_1 \leq 2\text{area}(S_{k-2}) \int_{0}^{\Delta} \phi_1^{k-2}d\phi_1 = \frac{2\text{area}(S_{k-2})}{k-1}\Delta^{k-1}.
\end{align*}

\medskip
\begin{lem}\label{lem:spherearearelation}
We have the following inequality
\[
\frac{(k-1)\text{area}(S_{k-1})}{\text{area}(S_{k-2})}\geq \pi.
\]
\end{lem}
Proof: It is easy to check the inequality for $k=1,\cdots, 5$. For $k\geq 6$, since $\text{area}(S_{k-1})= \frac{2\pi^{\frac{k}{2}}}{\Gamma(\frac{k}{2})}$ where $\Gamma(x)$ is the Gamma function, $\text{area}(S_{k})<\text{area}(S_{k-1})$ when $k> 5$. Therefore
\[
\frac{\text{area}(S_{k-1})}{\text{area}(S_{k-2})} > \frac{\text{area}(S_{k})}{\text{area}(S_{k-2})}= \frac{2\pi}{k-1}>\frac{\pi}{k-1} 
\] 
where the equality inside follows from a property of the Gamma function. This proves the lemma.

\begin{lem}\label{lem:highdnumberproject1}
For $n$ points $\vect{y}_j \in \mathbb R^k, k\geq 2, n\geq 2$, with minimum separation $d_{\min}:=\min_{p\neq j}||\vect{y}_p-\vect{y}_j||_2$, let $\Delta=\Big(\frac{\pi}{n(n-1)}\Big)^{\frac{1}{k-1}}$. There exists an unit vector $\vect{v}\in \mathbb R^k$ such that
\begin{equation}\label{equ:highdprojectionlower1}
\min_{p\neq j, 1\leq p, j \leq n}||\mathcal P_{\vect v^\perp}(\vect{y}_p)-\mathcal P_{\vect v^\perp}(\vect{y}_j)||_2\geq \frac{2\Delta d_{\min}}{\pi}. 
\end{equation}
\end{lem}
Proof: It is clear that there are at most $\frac{n(n-1)}{2}$ different $\vect u_{pj}= \vect y_p -\vect y_j, p<j$. Using Lemma \ref{lem:badareaestiamte} and \ref{lem:spherearearelation}, we have
\[
\text{area} \left(\cup_{p<j, 1\leq j, p\leq n} N(\vect{u}_{pj}, \Delta)\right) \leq \frac{n(n-1)}{2} \frac{2\text{area}(S_{k-2})}{k-1}\Delta^{k-1}\leq \text{area}(S_{k-1}).
\]
On the other hand, $\cup_{1\leq j<p\leq n} N(\vect{u}_{pj}, \Delta)$ is an open set in $S_{k-1}$. Thus $S_{k-1}\setminus \cup_{1\leq j< p\leq n} N(\vect{u}_{pj}, \Delta)$ is not empty. By the definition of $N(\vect{u}_{pj}, \Delta)$, there exists an unit vector $\vect{v} \in \mathbb R^{k}$ such that 
\[
||\mathcal P_{\vect v^\perp}(\vect u_{pj})||_2\geq d_{\min} \sin\Delta, \quad \forall 1\leq p<j\leq n. 
\]
Finally using the inequality $\sin \Delta \geq \frac{2}{\pi} \Delta$, we obtain (\ref{equ:highdprojectionlower1}) immediately.

\medskip
\begin{lem}\label{lem:highdmusicdirection}
For $n$ points $\vect{y}_j \in \mathbb R^2, n\geq 2$, with minimum separation $d_{\min}:=\min_{p\neq j}||\vect{y}_p-\vect{y}_j||_2$, let $\Delta=\frac{\pi}{n(n+1)}$ and choose $n$ unit vectors $\vect v_q=(\cos(2q\Delta), \sin(2q\Delta))^T\in \mathbb R^2, q=1,\cdots,\frac{n(n+1)}{2}$. There exists $q^*$ so that 
\[
\min_{p\neq j, 1\leq p, j \leq n}||\mathcal P_{v_{q^*}^\perp}(\vect{y}_p)-\mathcal P_{\vect v_{q^*}^\perp}(\vect{y}_j)||_2\geq \frac{2 d_{\min}}{n(n+1)}.
\]
\end{lem}
Proof: It is clear there are at most $\frac{n(n-1)}{2}$ different $\vect u_{pj}= \vect y_p -\vect y_j, 1\leq p<j\leq n$. Denote $\vect v(\theta) = (\cos \theta, \sin \theta)^T$, we observe that if $||\mathcal P_{\vect v(\theta)^\perp}(\vect{u})||_2 < ||\vect{u}||_2\sin 
\Delta, \theta\in [2\Delta,\pi]$, then $||\mathcal P_{\vect v(\theta^*)^\perp}(\vect{u})||_2 \geq ||\vect{u}||_2\sin 
\Delta$, for $|\theta^*-\theta|\geq 2\Delta, \theta^*\in [2\Delta,\pi]$. Therefore, recall definition (\ref{equ:defineofN}), if $\vect v_{q_0}\in N(\vect u_{p_0 j_0}, \Delta)$ for some $1\leq p_0, j_0\leq n$, then $\vect v_{q}\not \in N(\vect u_{p_0 j_0}, \Delta), \forall q\neq q_0, q=1,\cdots, \frac{n(n+1)}{2}$. Since we have $\frac{n(n+1)}{2}$ different $q$'s, there must be some $\vect v_{q^*} \not \in \cup_{p<j, 1\leq j, p\leq n}N(\vect u_{pj}, \Delta)$. Hence, 
\[
\min_{p\neq j, 1\leq p, j \leq n}||\mathcal P_{\vect v_{q^*}^\perp}(\vect{y}_p)-\mathcal P_{\vect v_{q^*}^\perp}(\vect{y}_j)||_2\geq d_{\min} \sin\Delta \geq d_{\min} \frac{2\Delta}{\pi},
\]
whence the lemma follows.

\vspace{0.5cm}
We next present a lemmas that is used to prove the results for support recovery. We consider the unit sphere in $\mathbb R^k$ and the following spherical coordinate 
\begin{equation}\label{equ:sphericalcoordinates1}
	\begin{aligned}
		x_{1}(\Phi)&=\cos(\phi_{1})\\
		x_{2}(\Phi)&=\sin(\phi_{1})\cos(\phi_{2})\\
		&\vdots \\
		x_{k-1}(\Phi)&=\sin(\phi_{1})\cdots \sin(\phi_{k-2})\cos(\phi_{k-1})\\
		x_{k}(\Phi)&=\sin(\phi_{1})\cdots \sin(\phi_{k-2})\sin(\phi_{k-1}),
	\end{aligned}
\end{equation}
where $\Phi=(\phi_1, \cdots , \phi_{k-1}) \in [0,\pi]^{k-2}\times [0, 2\pi)$. For $0<\theta<\frac{\pi}{2}$ and $N=\lfloor \frac{\pi}{2\theta} \rfloor$, we denote  
\begin{align}\label{equ:vectorlist1}
	\vect v_{\tau_1\cdots \tau_{k-1}}= \big(x_1(\Phi_{\tau_1\cdots\tau_{k-1}}), \cdots , x_k(\Phi_{\tau_1\cdots\tau_{k-1}})\big)^T, \quad 1\leq \tau_j \leq N,
\end{align}
where $\Phi_{\tau_1\cdots\tau_{k-1}}= (\tau_1\theta, \cdots, \tau_{k-1}\theta)$. It is obvious that $\Phi_{\tau_1\cdots\tau_{k-1}} \in [0, \frac{\pi}{2}]^{k-1}$ and $\vect v_{\tau_1\cdots \tau_{k-1}} \neq \vect v_{p_1\cdots p_{k-1}}$ if $(\tau_1, \cdots, \tau_{k-1})\neq (p_1, \cdots, p_{k-1})$. There are $N^{k-1}$ different unit vectors of the form (\ref{equ:vectorlist1}). 


\begin{lem}\label{lem:projectvectorangle1}
For two different vectors $\vect v_{\tau_1\cdots \tau_{k-1}} \neq \vect v_{p_1\cdots p_{k-1}}$ in (\ref{equ:vectorlist1}), we have
\begin{equation}\label{equ:projectlemangle1}
0 \leq \vect v_{\tau_1\cdots \tau_{k-1}} \cdot \vect v_{p_1\cdots p_{k-1}} \leq \cos\theta.
\end{equation}
\end{lem}
Proof: Because $\sin \tau_j \theta \geq 0, \cos \tau_j\theta \geq 0$, for $\tau_j$ in (\ref{equ:vectorlist1}), the first inequality, $0\leq \vect v_{\tau_1\cdots \tau_{k-1}} \cdot \vect v_{p_1\cdots p_{k-1}}$, is easy to verify using (\ref{equ:sphericalcoordinates1}). We next prove $\vect v_{\tau_1\cdots \tau_{k-1}} \cdot \vect v_{p_1\cdots p_{k-1}} \leq \cos\theta$ by induction. When $k=2$, it is clear that (\ref{equ:projectlemangle1}) holds. Suppose when $k = j-1$, (\ref{equ:projectlemangle1}) holds. For $k=j$, we consider the inner product between the vector $\vect v_{\tau_1\cdots \tau_{j-2}\tau_{j-1}}$ and  $\vect v_{p_1\cdots p_{j-2}p_{j-1}}$. We first observe that (using (\ref{equ:sphericalcoordinates1})), we have  
\[
\vect v_{\tau_1\cdots \tau_{j-2}\tau_{j-1}}\cdot \vect v_{p_1\cdots p_{j-2}\tau_{j-1}} = \vect v_{\tau_1\cdots \tau_{j-2}}\cdot \vect v_{p_1\cdots p_{j-2}}, 
\] 
where $\vect v_{p_1\cdots p_{j-2}\tau_{j-1}}$ is an unit vector in $\mathbb R^{j}$ and $\vect v_{\tau_1\cdots \tau_{j-2}}, \vect v_{p_1\cdots p_{j-2}}$ a unit vector in $\mathbb R^{j-1}$. By the assumption that (\ref{equ:projectlemangle1}) holds for $k=j-1$, the above equality gives 
\begin{align}\label{equ:projectlemangle2}
	0\leq \vect v_{\tau_1\cdots \tau_{j-2}\tau_{j-1}}\cdot \vect v_{p_1\cdots p_{j-2}\tau_{j-1}} \leq \cos \theta.
\end{align}
We then prove $ \vect v_{\tau_1\cdots \tau_{j-2}\tau_{j-1}}\cdot \vect v_{p_1\cdots p_{j-2}p_{j-1}}\leq \cos \theta$. Using the decomposition
\begin{align*}
\vect v_{p_1\cdots p_{j-2}p_{j-1}} = \vect v_{p_1\cdots p_{j-2}\tau_{j-1}} +  \vect x,
\end{align*}
where 
\begin{equation*}
\vect x = \sin(\tau_{1}\theta)\sin(p_{1}\theta)\cdots \sin(\tau_{j-2}\theta)\sin(p_{j-2}\theta)\big(0,\cdots, 0, \cos(p_{j-1}\theta)-\cos(\tau_{j-1}\theta),  \sin(p_{j-1}\theta)-\sin(\tau_{j-1}\theta)\big)^T,
\end{equation*}
we have
\begin{align*}
&\vect v_{\tau_1\cdots \tau_{j-1}}\cdot \vect v_{p_1\cdots p_{j-1}}= \vect v_{\tau_1\cdots \tau_{j-2}\tau_{j-1}} \cdot (\vect v_{p_1\cdots p_{j-2}\tau_{j-1}} +\vect x)\\
=&\vect v_{\tau_1\cdots \tau_{j-2}\tau_{j-1}} \cdot \vect v_{p_1\cdots p_{j-2}\tau_{j-1}} + \vect v_{\tau_1\cdots \tau_{j-2}\tau_{j-1}} \cdot \vect x \\
\leq& \cos \theta + \vect v_{\tau_1\cdots \tau_{j-2}\tau_{j-1}} \cdot \vect x \quad  (\text{by (\ref{equ:projectlemangle2})})\\
\leq& \cos \theta+\sin(\tau_{1}\theta)\sin(p_{1}\theta)\cdots \sin(\tau_{j-2}\theta)\sin(p_{j-2}\theta) \cos(\tau_{j-1}\theta)\big(\cos(p_{j-1}\theta)-\cos(\tau_{j-1}\theta)\big)\\
&+\sin(\tau_{1}\theta)\sin(p_{1}\theta)\cdots \sin(\tau_{j-2}\theta)\sin(p_{j-2}\theta) \sin(\tau_{j-1}\theta)\big(\sin(p_{j-1}\theta)-\sin({\tau_{j-1}}\theta)\big)\\
= &\cos\theta+ \sin(\tau_{1}\theta)\sin(p_{1}\theta)\cdots \sin(\tau_{j-2}\theta)\sin(p_{j-2}\theta) \big(\cos(\tau_{j-1}\theta)\cos(p_{j-1}\theta)+\sin(\tau_{j-1}\theta)\sin(p_{j-1}\theta)-1 \big)\\
\leq & \cos \theta. 
\end{align*}
This completes the induction argument and proves the lemma.

\medskip
\begin{lem}\label{lem:projectvectorangle2}
For a vector $\vect u\in \mathbb R^k$,  suppose $||\mathcal P_{\vect v_{\tau_1\cdots \tau_{k-1}}^\perp}(\vect u)||_2 < \sin (\frac{\theta}{2}) ||\vect u||_2$ with $\vect v_{\tau_1\cdots \tau_{k-1}}$ defined in (\ref{equ:vectorlist1}), we have $||\mathcal P_{\vect v_{p_1\cdots p_{k-1}}^\perp}(\vect u)||_2 \geq \sin (\frac{\theta}{2})||\vect u||_2$ for $\vect v_{p_1\cdots p_{k-1}} \neq \vect v_{\tau_1\cdots \tau_{k-1}}$.
\end{lem}
Proof: We denote the $2$-dimensional space spanned by $\vect v_{\tau_1\cdots \tau_{k-1}}$ and $\vect v_{p_1\cdots p_{k-1}}$ by $S$. We decompose $\vect u$ as $\vect u = \vect u_{S}+ \vect u_{S^{\perp}}$. The condition $||\mathcal P_{\vect v_{\tau_1\cdots \tau_{k-1}}^\perp}(\vect u)||_2 < \sin (\frac{\theta}{2})||\vect u||_2$ implies that
\[
||\mathcal P_{\vect v_{\tau_1\cdots \tau_{k-1}}^\perp}(\vect u)||_2^2 = ||\vect u_{S^\perp}||_2^2 + ||\vect u_{S}- \big(\vect u_S\cdot \vect v_{\tau_1\cdots \tau_{k-1}}\big) \vect v_{\tau_1\cdots \tau_{k-1}}||_2^2 < \sin(\frac{\theta}{2})^2||\vect u||_2^2.
\] 
Using the decomposition $\sin(\frac{\theta}{2})^2||\vect u||_2^2 = \sin(\frac{\theta}{2})^2(||\vect u_S||_2^2+||\vect u_{S^{\perp}}||_2^2)$, we further get 
\[||\vect u_{S}- \big(\vect u_S\cdot \vect v_{\tau_1\cdots \tau_{k-1}}\big) \vect v_{\tau_1\cdots \tau_{k-1}}||_2^2< \sin (\frac{\theta}{2})^2 ||\vect u_S||_2^2.
\] 
It follows that
\[
||\vect u_{S}||_2^2-|\vect u_S\cdot \vect v_{\tau_1\cdots \tau_{k-1}}|^2 = ||\vect u_{S}- \big(\vect u_S\cdot \vect v_{\tau_1\cdots \tau_{k-1}}\big) \vect v_{\tau_1\cdots \tau_{k-1}}||_2^2< \sin(\frac{\theta}{2})^2||\vect u_S||_2^2.
\]
Therefore, $|\vect u_S\cdot \vect v_{\tau_1\cdots \tau_{k-1}}|^2> \cos (\frac{\theta}{2})^2 ||\vect u_S||_2^2$. Denoting the angle between $\vect v_{\tau_1\cdots \tau_{k-1}}$ and $\vect u_S$ by $\angle(\vect v_{\tau_1\cdots \tau_{k-1}}, \vect u_S)$, we thus have $\angle(\vect v_{\tau_1\cdots \tau_{k-1}}, \vect u_S)<\frac{\theta}{2}$. On the hand, by Lemma \ref{lem:projectvectorangle1}, $0\leq \vect v_{\tau_1\cdots \tau_{k-1}}\cdot \vect v_{p_1\cdots p_{k-1}}\leq \cos \theta$ and this gives $\theta\leq \angle(\vect v_{\tau_1\cdots \tau_{k-1}}, \vect v_{p_1\cdots p_{k-1}}) \leq \frac{\pi}{2}$. Combining these estimates, we have
\begin{equation*}\label{equ:projectlemangle3}
\frac{\theta}{2}\leq \angle(\vect v_{p_1\cdots p_{k-1}}, \vect u_S) \leq \frac{\pi}{2}+\frac{\theta}{2}.    
\end{equation*}
Note that $\frac{\theta}{2}<\frac{\pi}{4}$, we can further get
\[
|\vect u_S \cdot \vect v_{p_1\cdots p_{k-1}}| \leq \cos(\frac{\theta}{2})||\vect u_S||_2.
\]
As a consequence, 
\[
||\mathcal P_{\vect v_{p_1\cdots p_{k-1}}^\perp}(\vect u)||_2^2 = ||\vect u_{S^\perp}||_2^2 + ||\vect u_{S}- \big(\vect u_S\cdot \vect v_{p_1\cdots p_{k-1}}\big) \vect v_{p_1\cdots p_{k-1}}||_2^2 \geq ||\vect u_{S^\perp}||_2^2 +   \sin(\frac{\theta}{2})^2||\vect u_S||_2^2 \geq \sin(\frac{\theta}{2})^2||\vect u||_2^2.
\]
This completes the proof.

\begin{lem}\label{lem:highdsupportproject1}
Let $\vect{y}_1, \vect{y}_2, \cdots, \vect{y}_n$ be $n$ different points in $\mathbb R^k$, $k \geq 2$.  Let ~$d_{\min}=\min_{p\neq j}||\vect{y}_p-\vect{y}_j||_2$ and $\Delta = \frac{\pi}{8}(\frac{2}{(n+2)(n-1)})^\frac{1}{k-1}$. Then there exist  $n+1$ unit vectors $\vect{v}_q$'s such that $0 \leq \vect v_p \cdot \vect v_j \leq \cos 2\Delta$ for $p\neq j$ and
\begin{equation}\label{equ:highdprojectionlower3}
\min_{p\neq j, 1\leq p, j \leq n}||\mathcal P_{\vect v_q^\perp}(\vect{y}_p)-\mathcal P_{\vect v_q^\perp}(\vect{y}_j)||_2\geq \frac{2\Delta d_{\min}}{\pi},\quad  q=1,\cdots, n+1. 
\end{equation}
\end{lem}
Proof: Note that there are at most $\frac{n(n-1)}{2}$ different vectors of the form $\vect u_{pj}= \vect y_p -\vect y_j, p<j$. For each $\vect u_{pj}$, consider the set $N(\vect{u}_{pj}, \Delta)$ defined in (\ref{equ:defineofN}). Let $\theta = 2\Delta$ and introduce the vectors $\vect v_{\tau_1\cdots \tau_{k-1}}$ as
in (\ref{equ:vectorlist1}). Using Lemma \ref{lem:projectvectorangle2}, we can derive that each set $N(\vect{u}_{pj}, \Delta)$ contains 
at most one of the vectors $\vect v_{\tau_1\cdots \tau_{k-1}}$'s. As a result, $\cup_{p<j, 1\leq j, p\leq n}N(\vect{u}_{pj}, \Delta)$ contains at most $\frac{n(n-1)}{2}$ vectors of the form $\vect v_{\tau_1\cdots \tau_{k-1}}$. 

Next recall that there are $N^{k-1}$ different vectors of the form in (\ref{equ:vectorlist1}), where $N = \lfloor \frac{\pi}{2\theta} \rfloor \geq \frac{\pi}{2\theta} -1 $. Since 
\[
\theta=2 \Delta =\frac{\pi}{4}(\frac{2}{(n+2)(n-1)})^\frac{1}{k-1}, 
\]
we have
\begin{align*}
N^{k-1}\geq \Big(\frac{\pi}{2\theta}-1\Big)^{k-1} = \Big(2 (\frac{(n+2)(n-1)}{2})^{\frac{1}{k-1}}-1\Big)^{k-1}
\geq \Big((\frac{(n+2)(n-1)}{2})^{\frac{1}{k-1}}\Big)^{k-1}= \frac{(n+2)(n-1)}{2}.
\end{align*}
Note that $\frac{(n+2)(n-1)}{2} - \frac{n(n-1)}{2} = n+1$, we can find $n+1$ vectors of the form $\vect v_{\tau_1\cdots \tau_{k-1}}$ that are not contained in the set  $\cup_{p<j, 1\leq j, p\leq n} N(\vect u_{pj}, \Delta)$. That is, we can find $n+1$ unit vectors, say, $\vect v_q$, $1\leq q \leq n+1$, which satisfy (\ref{equ:highdprojectionlower3}). Moreover, by Lemma \ref{lem:projectvectorangle1}, these vectors also satisfy the condition that $0 \leq \vect v_p \cdot \vect v_j \leq \cos 2\Delta$ for  $p\neq j$. This completes the proof.

\medskip
\begin{lem}\label{lem:highdsupportproject2}
Let $k\geq 2$. For a vector $\vect{u}\in \mathbb R^k$, and two unit vectors $\vect{v}_1, \vect{v}_2\in \mathbb R^k$ satisfying $0 \leq \vect v_1 \cdot \vect v_2 \leq \cos \theta$, we have 
\begin{equation}\label{equ:highdprojectionlower4}
||\mathcal P_{\vect v_1^{\perp}}(\vect u)||_2^2+||\mathcal P_{\vect v_2^{\perp}}(\vect u)||_2^2  \geq (1-\cos(\theta))||\vect u||_2^2.
\end{equation}
\end{lem}
Proof: We first prove the lemma for dimension two. Indeed, for $\vect{u}\in \mathbb R^2$, and two unit vectors $\vect{v}_1, \vect{v}_2\in \mathbb R^2$ satisfying $0 \leq \vect v_1 \cdot \vect v_2 \leq \cos \theta$, we have 
\begin{equation}\label{equ:projectionlower3}
\btwonorm{(\vect{v}_1\cdot \vect u,\  \vect{v}_2\cdot \vect u)^T}^2
= \btwonorm{\begin{pmatrix}
	\vect{v}_1^T\\
	\vect{v}_2^T
	\end{pmatrix}\cdot \vect u}^2 \geq \sigma_{\min}^2(\begin{pmatrix}
\vect{v}_1^T\\
\vect{v}_2^T
\end{pmatrix})||\vect u||_2^2 \geq (1-\cos \theta)||\vect u||_2^2,
\end{equation}
where the last inequality follows from calculating $\sigma_{\min}(\begin{pmatrix}
\vect{v}_1^T\\
\vect{v}_2^T
\end{pmatrix})$. We now prove the lemma for $k$ dimensional case. When $\cos \theta =1$, the lemma obviously holds. When $\cos \theta<1$, we denote the $2$-dimensional space spanned by $\vect v_1$ and $\vect v_2$ as $S$. Let $\vect g_1$ and $\vect g_2$ be the unit vectors in $S$ that are perpendicular to $\vect v_1$ and $\vect v_2$ respectively and satisfy $0\leq  \vect g_1 \cdot \vect g_2 \leq \cos \theta$ as well. Then $\vect u$ has the decomposition 
\[
\vect u = \mathcal P_{S^{\perp}}(\vect u)+ \lambda_j \vect g_j +  \eta_j\vect v_j, \quad j=1,2,
\]
where $ \lambda_j = \mathcal P_{S}(\vect u)\cdot \vect g_j$ and $ \eta_j = \mathcal P_{S}(\vect u)\cdot \vect v_j$. Since $S$ is a 2-dimensional subspace, applying (\ref{equ:projectionlower3}) in $S$, we have $||(\lambda_1, \lambda_2)^T||_2^2 =||(\mathcal P_{S}(\vect u)\cdot \vect g_1, \mathcal P_{S}(\vect u)\cdot \vect g_2)^T||_2^2 \geq (1-\cos \theta)||\mathcal P_{S}(\vect u)||_2^2$. Note that $\mathcal P_{\vect v_j^{\perp}}(\vect u) = \mathcal P_{S^\perp}(\vect u)+ \lambda_j \vect g_j, j=1,2$, we further have 
\begin{align*}
&||\mathcal P_{\vect v_1^{\perp}}(\vect u)||_2^2+||\mathcal P_{\vect v_2^{\perp}}(\vect u)||_2^2 = 2||\mathcal P_{S^\perp}(\vect u)||_2^2+ \lambda_1^2+\lambda_2^2 \\
\geq & 2||\mathcal P_{S^\perp}(\vect u)||_2^2+ (1-\cos\theta)||\mathcal P_{S}(\vect u)||_2^2
\geq (1-\cos\theta)||\vect u||_2^2.
\end{align*}
This completes the proof of the lemma.


\bibliographystyle{plain} 
\bibliography{references} 

\end{document}